\documentclass[onecolumn,12pt,amssymb,amsmath,preprintnumbers,nofootinbib,superscriptaddress,aps,preprintnumbers]{revtex4-1}
\usepackage{bm,color,xcolor}
\usepackage{slashed} 
\usepackage{graphicx}
\usepackage{mwe}
\usepackage{adjustbox}
\usepackage[compat=1.1.0]{tikz-feynman}
\usepackage{float}
\usepackage[utf8x]{inputenc}
\usepackage[T1]{fontenc}

\usepackage{multirow}
\usepackage{tablefootnote}
\usepackage{comment}
\usepackage{mathtools}
\usepackage{appendix}
\usepackage[colorlinks=true
,urlcolor=blue
,anchorcolor=blue
,citecolor=blue 
,filecolor=blue
,linkcolor=blue
,menucolor=blue
,linktocpage=true
,pdfproducer=medialab
,pdfa=true
]{hyperref}
\usepackage{fontawesome} 
\usepackage{bm,color,color,soul}
\definecolor{UMNRed}{RGB}{144, 0, 33}
\definecolor{UMNGold}{RGB}{255 204 51}

\DeclareMathOperator{\Br}{Br}

\newcommand{\beq}{\begin{equation}}
\newcommand{\eeq}{\end{equation}}
\newcommand{\bea}{\begin{eqnarray}}
\newcommand{\eea}{\end{eqnarray}}

\newcommand{\brr}[1]{\left(#1\right)}
\newcommand{\srr}[1]{\left[#1\right]}

\begin{document}
\renewcommand{\thefootnote}{\arabic{footnote}}
 
\begin{flushright}
\end{flushright}

\title{\bf  
Higgs Width and Couplings at High Energy Muon Colliders with Forward Muon Detection
}



\author{Peiran Li}
\thanks{li001800@umn.edu, \scriptsize \!\! \href{https://orcid.org/0009-0005-7748-7085}{0009-0005-7748-7085}}
\author{Zhen Liu}
\thanks{zliuphys@umn.edu, \scriptsize \!\! \href{https://orcid.org/0000-0002-3143-1976}{0000-0002-3143-1976}}
\author{Kun-Feng Lyu}
\thanks{lyu00145@umn.edu, \scriptsize \!\! \href{https://orcid.org/0000-0002-3291-1701}{0000-0002-3291-1701}}
 
\smallskip
\medskip

\affiliation{\it School of Physics and Astronomy, University of Minnesota, Minneapolis, MN 55455, USA}

\medskip

\begin{abstract}
\vspace{0.5cm}

We propose a novel method using the $ZZ$-fusion channel and forward muon detection at high-energy muon colliders to address the challenge of the Higgs coupling-width degeneracy. Our approach enables inclusive Higgs rate measurement to 0.75\% at 10~TeV muon collider, breaking the coupling-width degeneracy. Results indicate the potential to refine Higgs coupling to sub-percent levels and estimate its total width within (-0.41\%, +2.1\%). Key insights include the effectiveness of forward muon tagging in signal-background separation despite broad recoil mass distribution due to muon energy reconstruction and beam energy spread. The study emphasizes the significance of muon rapidity coverage up to $|\eta (\mu)|<6$, enhancing measurement precision. Our findings highlight the unique capabilities of high-energy lepton colliders for model-independent Higgs coupling determination and lay the groundwork for future advancements in muon collider technology and Higgs physics research.

\end{abstract}

\preprint{UMN-TH-4309/24}
\maketitle
\setcounter{tocdepth}{1}
\tableofcontents


\newpage

\section{Introduction}

The core quest of the next collider is to study the Higgs boson properties and reveal the many profound fundamental puzzles around this new boson~\cite{deBlas:2019rxi,EuropeanStrategyforParticlePhysicsPreparatoryGroup:2019qin,Dawson:2022zbb,Narain:2022qud}. Many future colliders are proposed, and their physics potentials are thoroughly understood, such as CEPC~\cite{An:2018dwb,CEPCStudyGroup:2018ghi,CEPCPhysicsStudyGroup:2022uwl}, ILC~\cite{Asner:2013psa,ILCInternationalDevelopmentTeam:2022izu}, FCC-ee~\cite{Benedikt:2651299,Bernardi:2022hny}, CLIC~\cite{CLICdp:2018cto} and C3~\cite{Vernieri:2022fae}. 
One key question is to pin down the Higgs overall coupling scale, which coincides with the Higgs width measurement. Besides a high signal background ratio that helps measure Higgs property, these future $e^+e^-$ colliders measure the overall coupling strength through an inclusive Higgs measurement in the $e^+e^-\rightarrow ZH$ process, without dependence on the Higgs exclusive decay modes, reconstructed using the recoil mass derived from the associated $Z$ boson four-momentum. 

Future high-energy muon collider generates vigorous interest and promises due to their direct access to ten-TeV scale physics. The high-energy muon collider can also copiously produce the Higgs boson mainly through the vector boson fusion (VBF) process, dominated by $WW$-fusion. We generally understand its Higgs performance at various stages~\cite{Buttazzo:2020uzc,Han:2020pif,Han:2021lnp,Forslund:2022xjq,Black:2022cth,Chen:2022yiu,Ruhdorfer:2023uea,Liu:2023yrb,Forslund:2023reu,Han:2023njx,Celada:2023oji}. However, high energy muon collider alone cannot pin down the Higgs coupling scale as no inclusive Higgs measurement anticipated. In this work, we point out that studying the $ZZ$-fusion channel with forward muon tagging would enable this hallmark of Higgs property measurement, namely, the absolute scale of its coupling and its width. 

We quantify many essential aspects to establish this channel and show its critical impact on the Higgs width, coupling determination, and global fit. The paper is organized as follows. In \autoref{sec:ZZ_inclusive}, we first analyze the signal and background and then make a simulation so that the precision on $h Z Z$ coupling measurement can be extracted. Then, in \autoref{sec:global_fit}, we develop global fits combined with other channels and show the sensitivity on various Higgs couplings. We summarize the key of this study in \autoref{sec:conclusion}.

\section{$ZZ$-Fusion for the Inclusive Rate} 
\label{sec:ZZ_inclusive}

For high-energy lepton colliders, the dominant single Higgs production channel is from the VBF process, especially the $WW$ fusion channel. At the high energy, the parton distribution function of gauge boson~\cite{Han:2020uid,Han:2021kes,Kane:1984bb,Dawson:1984gx,Fornal:2018znf,Chen:2016wkt,Ruiz:2021tdt} split from the incoming muons can be significantly enhanced.
For the $WW$ fusion, the associated particles to the Higgs boson are the invisible neutrinos emitted in the forward direction. In the $ZZ$ fusion, the particles associated with the Higgs boson are the forward muons. We can separate the two primary channels if one can tag the forward muons.
To quantify the projected sensitivity of this channel, we use the 10 TeV muon collider benchmark with an integrated luminosity of $10~\text{ab}^{-1}$. We also show the distribution and analysis results for a 3~TeV muon collider, which can be an essential stage. 
The $WW$-fusion, $\mu^+ \mu^- \rightarrow \bar{\nu}_\mu \nu_\mu h$, has signal rate close to 1~pb. The $ZZ$-fusion of $\mu^+ \mu^- \rightarrow \mu^+ \mu^- h$ follows, with around one order smaller cross-section. The $WW$ fusion leads to the invisible neutrinos, while the $ZZ$ fusion is accompanied by the potentially detectable muon in the forward direction. If the forward muons can be tagged by placing the detector in the high $\eta$ region, one can exploit the information of such muons.

In this study, we quantify the behavior of forward muons and the physics gain in Higgs precision from such detection. The recoil mass is defined by
\begin{equation}
    \text{M}^2_{\rm recoil} = \srr{ p_1(\mu^+) + p_2(\mu^-) - k_1(\mu^+) - k_2(\mu^-) }^2
\end{equation}
where $p_1$ and $p_2$ are the momenta of the incoming muons, $k_1$ and $k_2$ refer to the momenta of outgoing forward muons. Such dimuon quantity could help us extract the inclusive scattering information of neutral gauge boson scattering, regardless of the final states from the scattering. 

\begin{figure}[th]
    \includegraphics[width=0.7\textwidth]{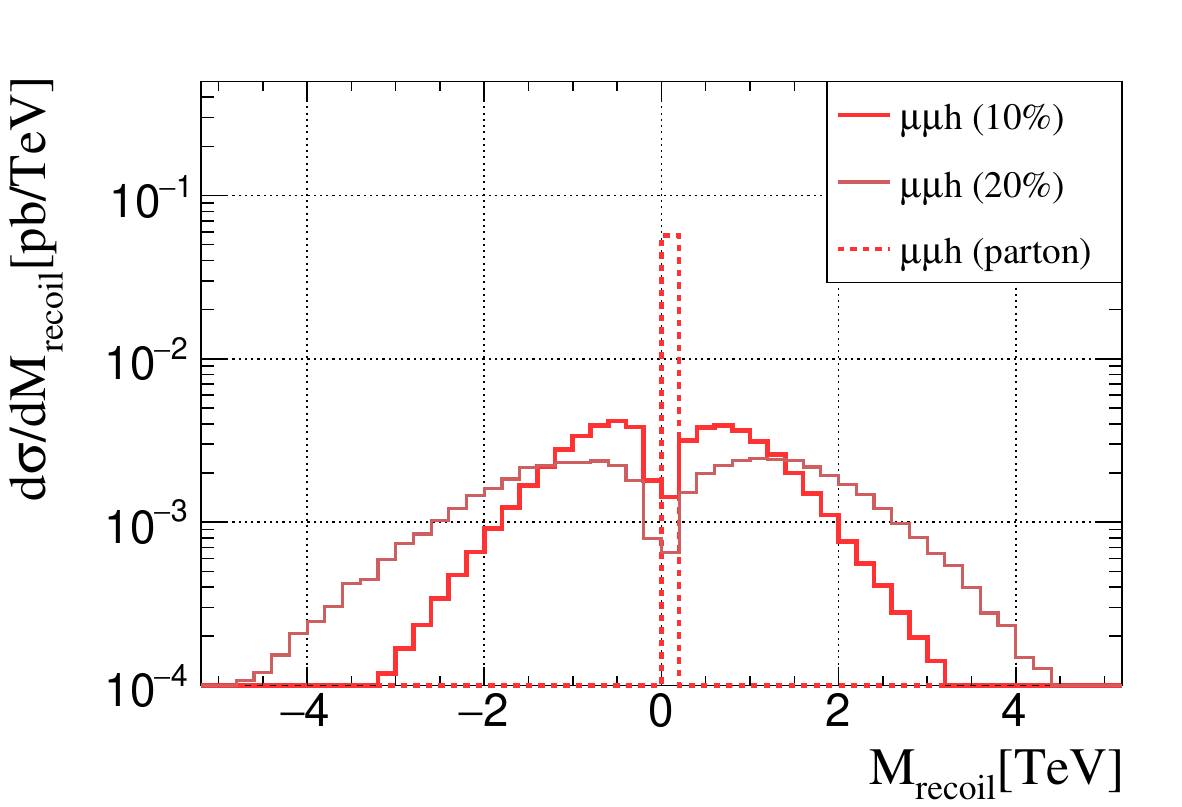}
    \caption{The recoil mass distributions of parton level (dashed) and reconstruction level (solid) at a 10~TeV muon collider. Two solid lines represent 10\% and 20\%  momentum resolution at the reconstruction level. The negative value of recoil mass $\text{M}_\text{recoil}$ is defined when $\text{M}_\text{recoil}^2<0$. At the parton level, the signal should be a Breit-Wigner peak center at the Higgs mass of 0.125~TeV.
    }
    \label{fig:recoil_parton_reconstruction}
\end{figure}

In \autoref{fig:recoil_parton_reconstruction}, we show the recoil mass distribution at the parton level (dashed line) and after reconstruction (solid line) under two different momenta resolutions; the details are provided in later sections. Ideally, parton-level events have a sharp recoil mass peak at the Higgs mass pole. However, energy resolution for the high energy and forward muons erodes this information to be much less powerful. One also omits the information that is further reconstructable from the scattering events, which requires careful determination of the observational possibility. We first classify the signal and background. The parton level events are generated using {\tt MadGraph5\_aMC}~\cite{Alwall:2014hca}. We implement a parton-level {\it pre-selection}\footnote{to avoid IR singularities, and as detector reconstruction-level will be more complicated than those} when generating both signal and background samples: 
\[p_T(\ell,j)>5~\text{GeV},~p_T(\gamma)>1~\text{GeV},~0<\eta(\ell)<10,~\Delta R(jj,j\ell,\ell\ell)>0.2.\]
Then the parton level events are passed to the interfaced {\tt Pythia}~\cite{Bierlich:2022pfr} and {\tt Delphes}~\cite{deFavereau:2013fsa,Mertens:2015kba} to simulate the reconstructed events we measure from the (forward) detectors.

\subsection{Signal and Background Considerations}

\begin{figure}[th]
    \includegraphics[width=0.495\textwidth]{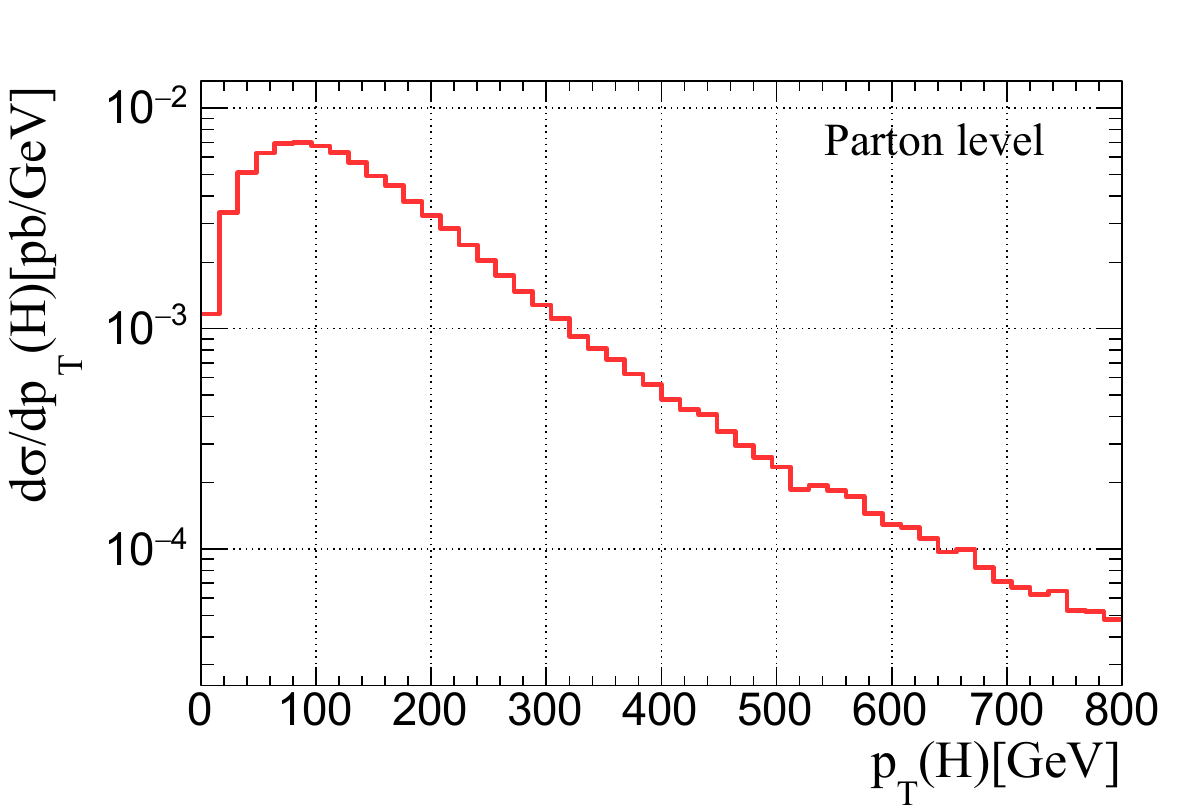}
    \includegraphics[width=0.495\textwidth]{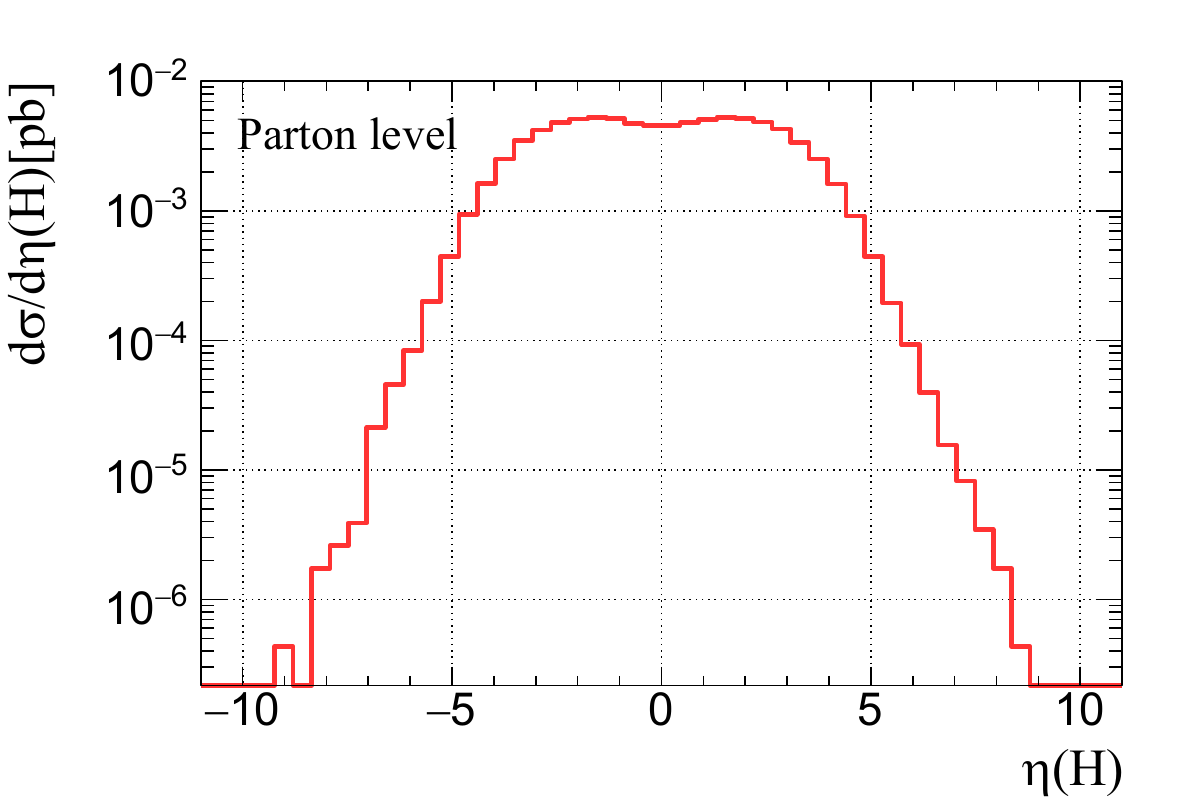}
    \includegraphics[width=0.495\textwidth]{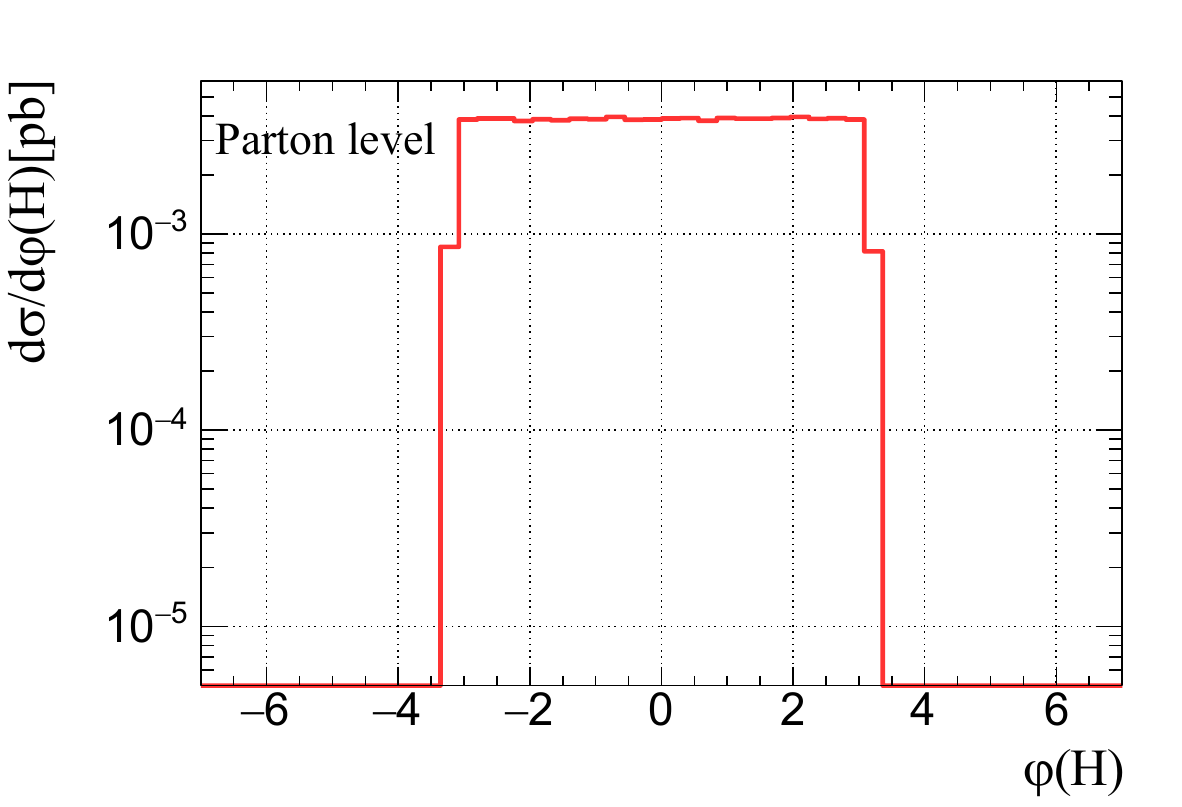}
    \caption{The parton level distributions of Higgs kinematics after parton-level pre-selection.
    }
    \label{fig:Higgs_parton}
\end{figure}

The signal is the $ZZ$-fusion Higgs production, $\mu^+ \mu^- \rightarrow \mu^+ \mu^- h$. Under the forward muons requirement (included in the pre-selection at the parton level)
the primary contribution is $ZZ$-fusion process with the cross-section around 86.7~fb as shown in \autoref{Table:parton_level_xs}. The parton-level kinematics of the Higgs boson are shown in \autoref{fig:Higgs_parton}. Due to the unpolarized initial muon beams, the produced Higgs boson is symmetric in $\eta$ and $\phi$. We can also see that the Higgs boson $p_T$ is dominant at around 100~GeV, the typical energy scale of the production, and spreads up to TeV. %

\begin{figure}[th]
    \centering 
    \includegraphics[width=0.49\textwidth]{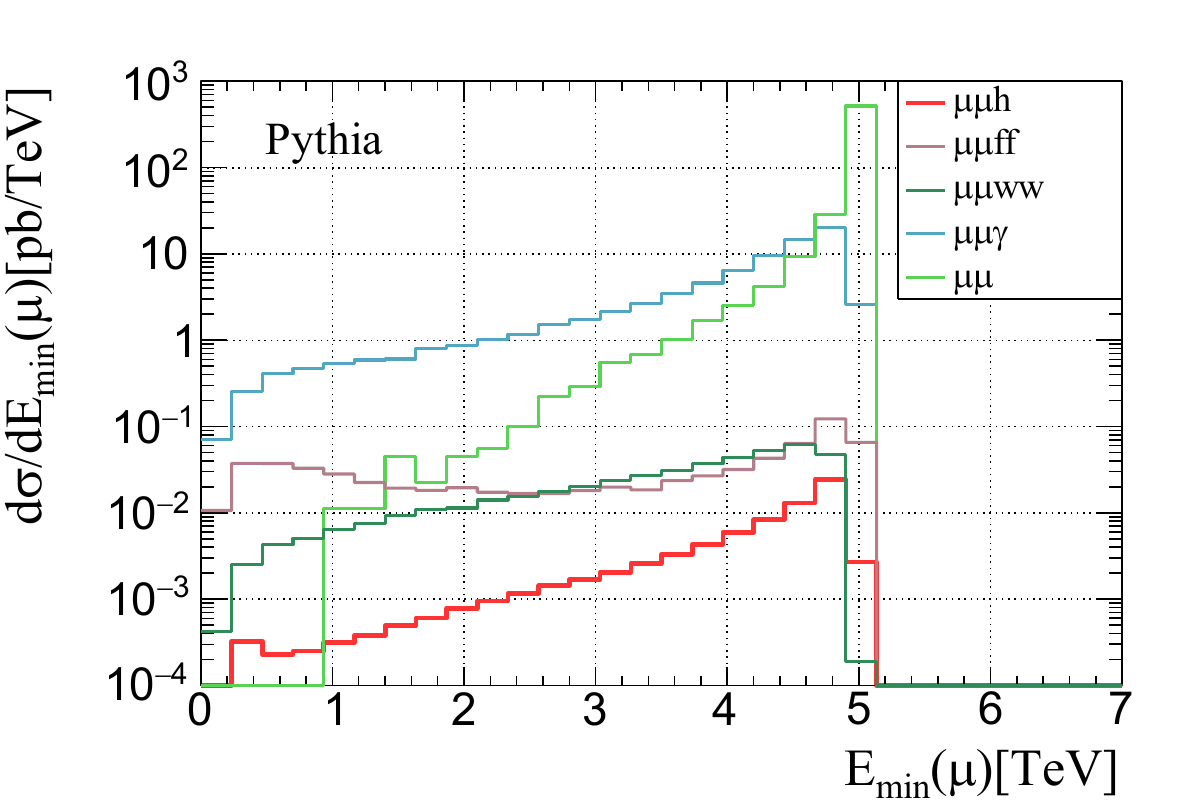}
    \includegraphics[width=0.49\textwidth]{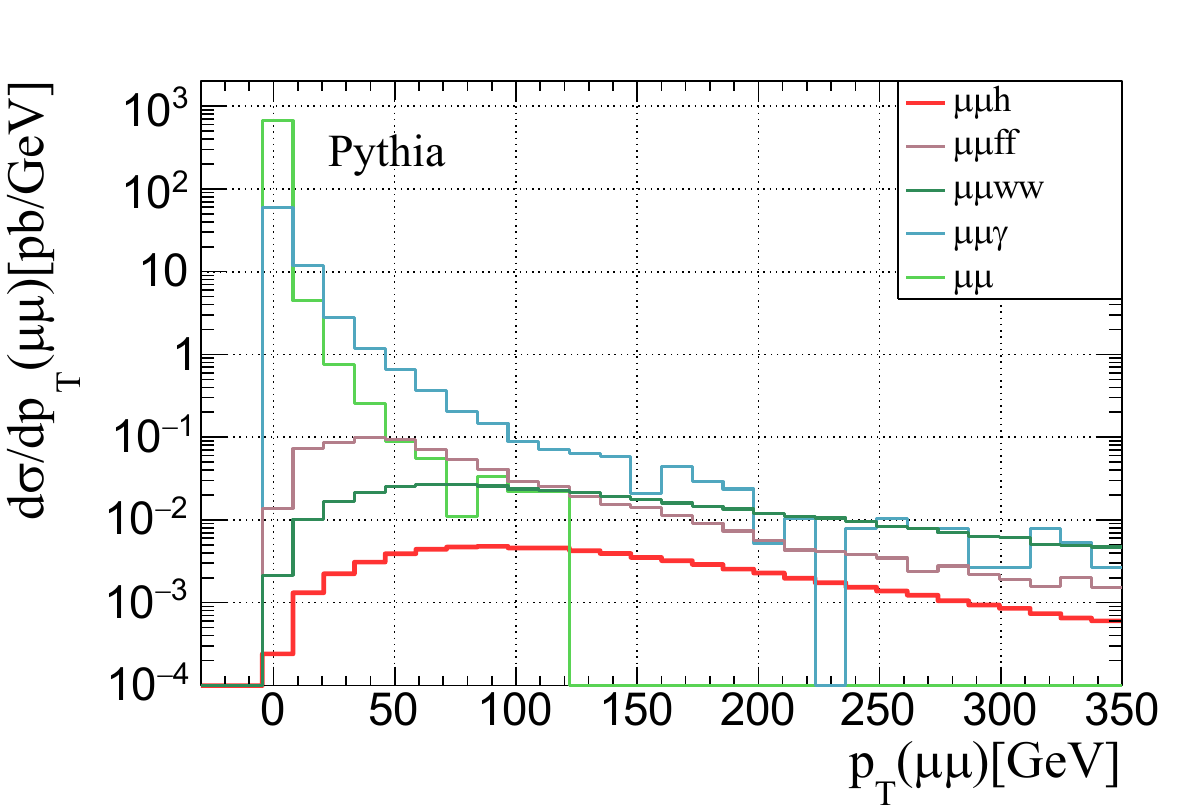}
    \vspace{1ex}
    \includegraphics[width=0.49\textwidth]{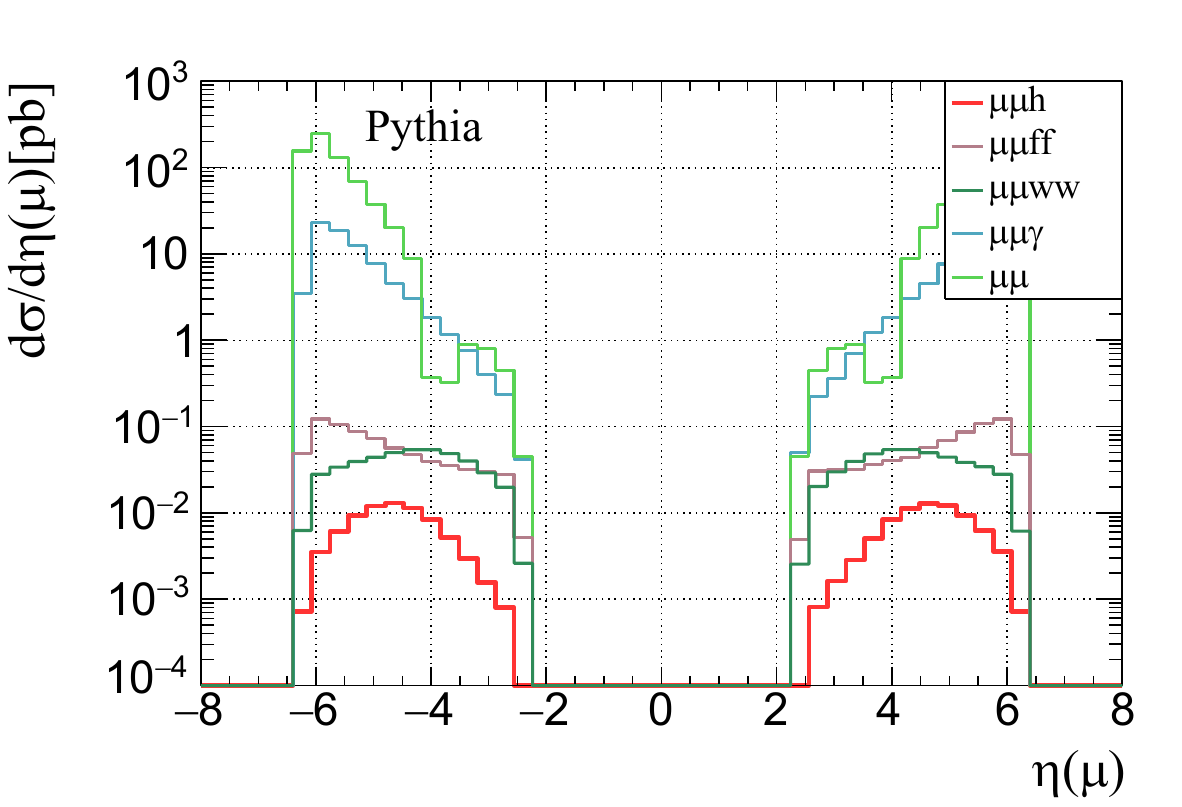}
    \includegraphics[width=0.49\textwidth]{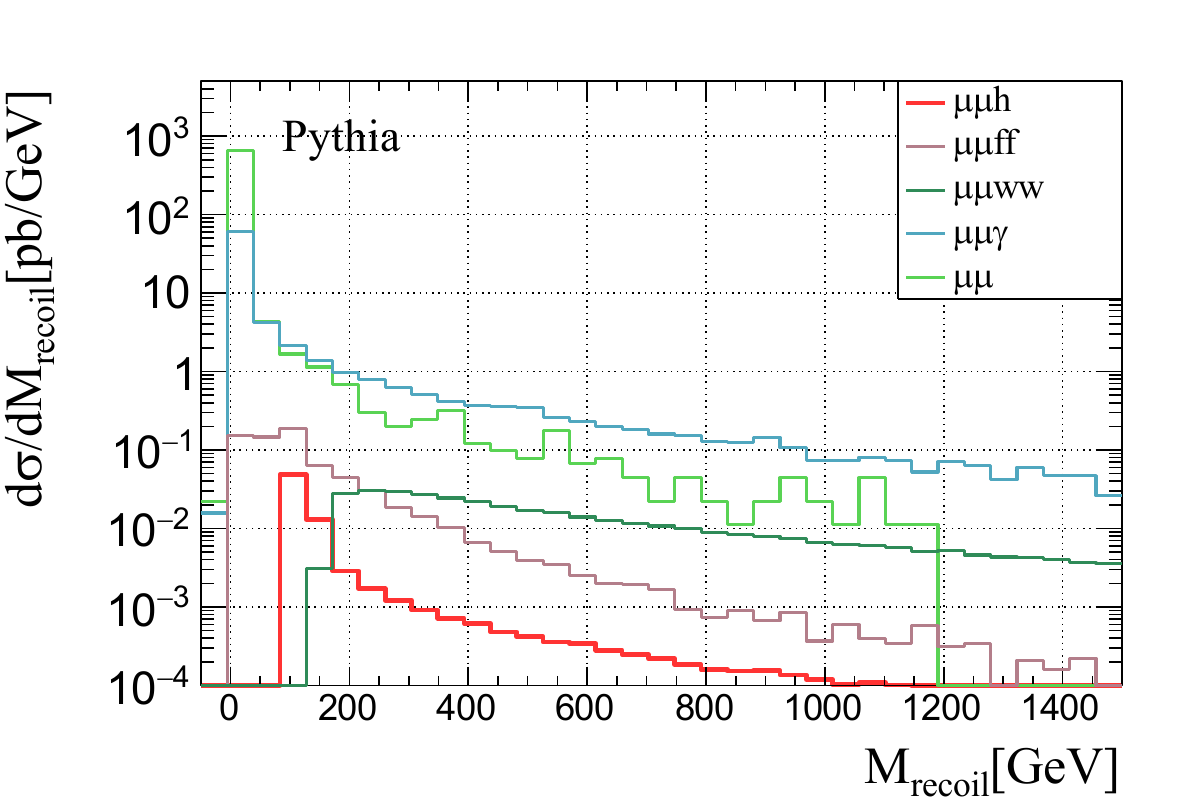}
    \caption{Truth level (including parton shower) muon kinematics distributions with cuts of $2.5<|\eta(\mu)|<8$, $\eta(\mu^-)\cdot\eta(\mu^+)<0$, and $p_T(\mu)>20~$GeV.} 
    \label{fig:kinematics_truth_level}
\end{figure}

The kinematics of muon is the main object in this study, and its truth level (including QED shower) information is shown as the red histogram in \autoref{fig:kinematics_truth_level}. We display the differential distribution over the minimum energy of the muon $E_\text{min}(\mu)$, the dimuon transverse momentum sum $p_T(\mu\mu)$, the pseudorapidity of each muon $\eta(\mu)$ and the recoil mass $\text{M}_\text{recoil}$. The primary change is that the parton shower spreads the recoil mass distribution with a long tail instead of the delta-like peak at $m_h$.

\begin{table}[th]
\centering
\begin{tabular}{|c|c|c|}
\hline
Type & Scattering process & cross-section $\sigma$ (pb) \\
\hline
VBF & $\mu^+ \mu^- \to \mu^+ \mu^- h$ & $8.67\times 10^{-2}$  \\
\hline
$t$-channel & $\mu^+ \mu^- \to \mu^+ \mu^-$ & $1.12\times 10^4$ \\
\hline
$t$-channel & $\mu^+ \mu^- \to \mu^+ \mu^- \gamma$ & $7.55 \times 10^2$ \\
\hline
VBS & $\mu^+ \mu^- \to \mu^+ \mu^- \ell^+\ell^-$ & 3.96  \\
\hline
VBS & $\mu^+ \mu^- \to \mu^+ \mu^- jj$ & 2.06  \\
\hline
VBS & $\mu^+ \mu^- \to \mu^+ \mu^- \nu_\ell \Bar{\nu}_\ell$ & 1.68  \\
\hline 
VBS & $\mu^+ \mu^- \to \mu^+ \mu^- W^+ W^-$ & $0.939$  \\
\hline
\end{tabular}%
\caption{Cross-section for both signal and background after parton-level pre-selection.
}
\label{Table:parton_level_xs}
\end{table}

The background events include $\mu^+ \mu^- \rightarrow \mu^+ \mu^- + X $ with $X$ to be any particles except the Higgs boson. The major background channels are listed in the following:
\begin{enumerate}
    \item $\mu^+ \mu^- \rightarrow \mu^+ \mu^-$ and $\mu^+ \mu^- \rightarrow \mu^+ \mu^- \gamma$
    \item $\mu^+ \mu^- \rightarrow \mu^+ \mu^-  f \bar{f}$ ($ f= l ,\nu, j$) 
    \item  $\mu^+ \mu^- \rightarrow \mu^+ \mu^- W^+ W^-$.
\end{enumerate}
Their cross-section and truth level (including parton shower) distributions for both the signal and these backgrounds are shown in \autoref{Table:parton_level_xs} and \autoref{fig:kinematics_truth_level}.

The first class is the elastic Bhabha scattering and those with one (visible) photon emission process. At the parton level, this is a 2-to-2 scattering process whose peak is at the ultra-forward region due to the $t$-channel enhancement. However, this is only a tree-level description. Furthermore, one must consider the real photon emission in the forward region plus the higher-order QED correction to cancel the infrared (IR) and collinear divergence. One feasible way to handle this is to merge the Parton shower sample events of the Bhabha scattering with the process with one additional photon $\mu^+ \mu^- \rightarrow \mu^+ \mu^- \gamma$, which will be elaborated afterward. We ignore more photon events due to the QED suppression. 

The second class of background is the forward muon pairs plus SM fermion pairs. The Feynman diagrams can be further divided into two categories. 
The forward muon scattering could have initial- and final-state radiations of (on- or off-shell) SM gauge bosons that further split into SM fermions. 
The other category is the neutral VBS process that produces fermion pairs through $s$-channel or $t$-channel process. 

The final states for the third class are $\mu^+\mu^-\to\mu^+\mu^-+ff\Bar{f}\Bar{f}$ where the four fermions do not need to be the same flavor and are mainly from the on-shell $W$ bosons decay. The contribution from on-shell $Z$ boson decay is much smaller. The lower rate is caused by the suppressed ``hard'' scattering cross-section for $Z Z \rightarrow Z Z$ in which only intermediate Higgs boson can be exchanged. While for $W^+ W^-$ boson final states, both gamma and $Z$ boson can serve as initial states, leading to a much higher cross-section. It turns out that the cross-section of $\mu^+\mu^-\to\mu^+\mu^- Z Z$ is much smaller than $\mu^+\mu^-\to\mu^+\mu^-W^+W^-$ by a factor of $10^4$ approximately hence we ignore it.

\begin{figure}[H]
    \centering
    \includegraphics[width=7.5cm]{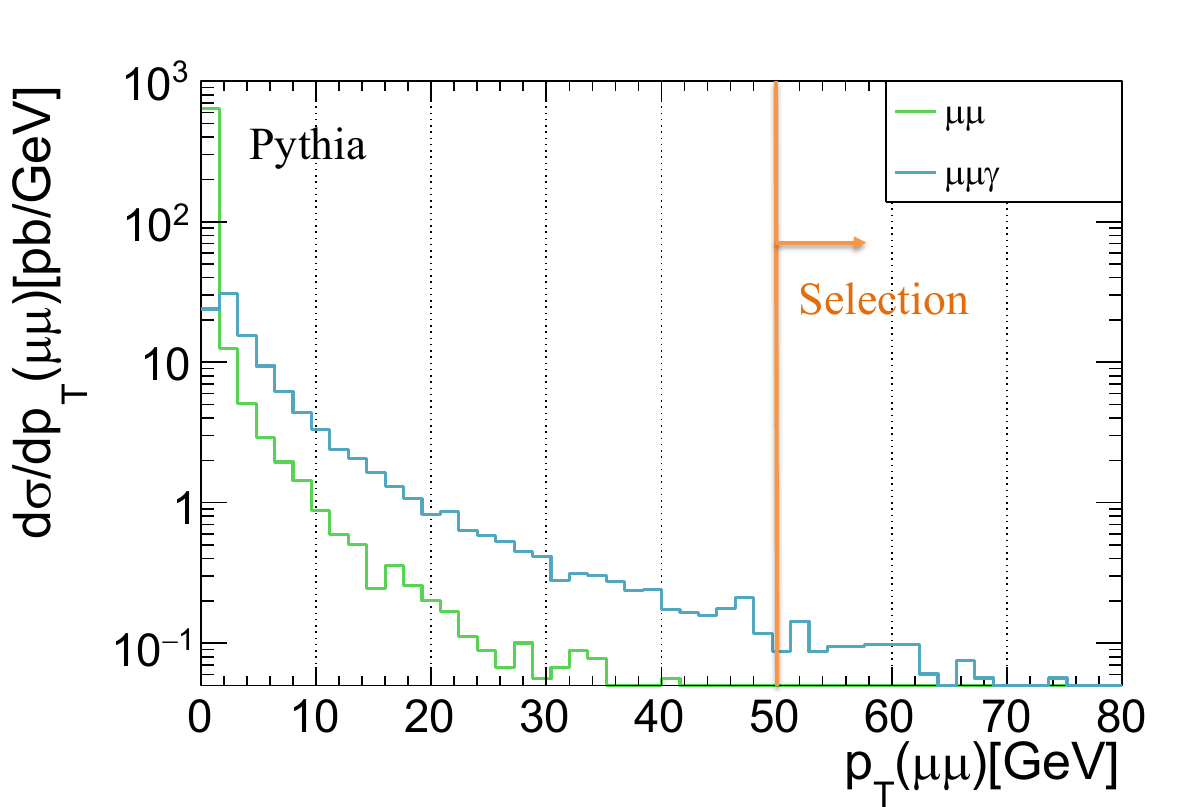}
    \qquad
    \includegraphics[width=7.5cm]{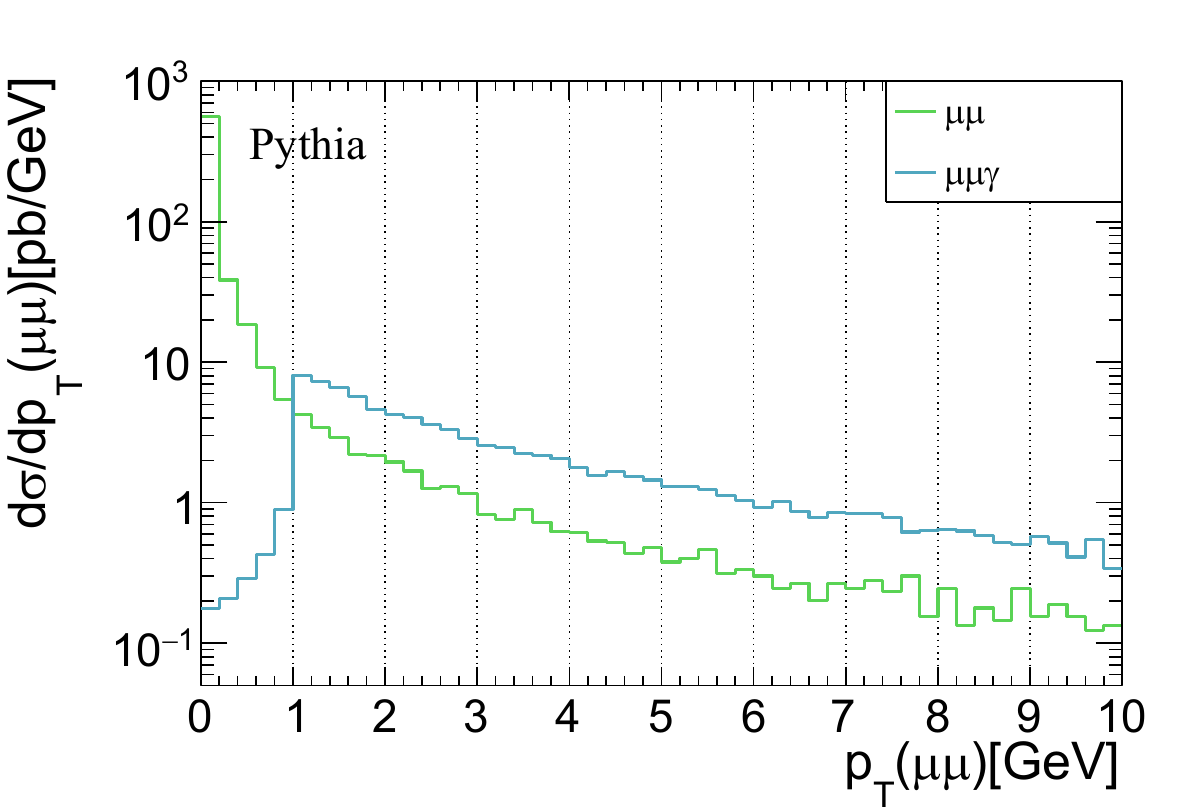}
    \caption{The truth level $p_T(\mu\mu)$ distribution after QED parton shower, under the cuts of $2.5<|\eta(\mu)|<8$, $\eta(\mu^+)\eta(\mu^-)<0$ and $p_T(\mu)>20~$GeV. The right panel zooms in on the left plot under $10~$GeV. The sharp cliff around 1 GeV is caused by a $p_T(\gamma)>1~$GeV cut.}
    \label{fig:sumPT_mumu_mumua_truth}
\end{figure}

All backgrounds are generated via MadGraph and Pythia. 
The primary background $\mu^+ \mu^- \rightarrow \mu^+ \mu^- (\gamma)$, whose event number is around 4 or 5 orders larger, needs to be carefully analyzed. The matrix element is reliable for the larger transverse momentum $p_T$ of $\mu^+ \mu^-$ pair or the photon. For a lower $p_T$ value, one has to consider the Infrared (IR) effects in which resummation is needed. We turn on the QED parton shower for both Bhabha scattering and $\mu^+ \mu^- \to \mu^+ \mu^- \gamma$, and include Initial State Radiation (ISR) and Final State Radiation (FSR). Furthermore, double counting must be avoided due to an overlap of phase space (final states) through matching and event merging. When the photon of a $\mu^+ \mu^- \to \mu^+ \mu^- \gamma$ process is soft/collinear, it is already counted inside the final states of Bhabha scattering after the QED shower. Similarly, a Bhabha scattering process, whose final state has a hard photon coming out after QED shower, is included in the phase space of $\mu^+ \mu^- \to \mu^+ \mu^- \gamma$. Therefore, when plotting the distribution of $p_T(\mu\mu)$, we expect there is a matching scale where both their cross-section and trend are the same. This feature is checked in truth level as shown in \autoref{fig:sumPT_mumu_mumua_truth}. By observing a matching scale around $0.8$ GeV, the lower $p_T$ region is dominated by Bhabha scattering, and the higher $p_T$ region should be contributed by $\mu^+ \mu^- \to \mu^+ \mu^- \gamma$. The $\mu^+ \mu^- \gamma$ process has a cliff at $1$ GeV due to the parton level cut $p_T(
\gamma)>1$ GeV. We will apply a cut of $p_T(\mu\mu)>50$ GeV at the detector level so that only $\mu^+ \mu^- \to \mu^+ \mu^- \gamma$ process survive. Given the large separation of scales we make, we can safely ignore the theoretical uncertainty on the pure Bhabha backgrounds.

\subsection{Forward Muon Tagging}
\label{sec:forward_detector}

As an unstable particle, muon colliders have a unique source of background which is from the decay products and interactions of them, called Beam Induced Background (BIB)~\cite{Ally:2022rgk,MuonCollider:2022glg,MuonCollider:2022ded}. To ensure the delivery of high-precision central physics, the muon collider detector designs have two tungsten cone-shaped shields (nozzles) around the beampipe. This shield limits the available angular coverage of the detected interaction region to $\theta > 10^\circ$, corresponding to $\eta < 2.4$.

The tungsten nozzles are initially designed for $\sqrt{s}=1.5~$TeV to shield beam-induced background. For a 10 TeV muon collider, sharper-cone nozzles can be further optimized. The potential of forward muon detectors has already been highlighted in several muon collider reviews~\cite{AlAli:2021let,Accettura:2023ked}. TeV muons can penetrate the shielding nozzle. Then, tagging and measuring forward muon with momentum can be exploited by adding a tracking station in the forward region, such as the design of FCC-hh~\cite{FCC:2018vvp}. We note here that the forward muon tagging does not require high precision in energy reconstruction. Hence, the energy loss for muons transpassing the tungsten nozzles does not affect the physics performance for the Higgs physics discussed here.

\subsection{Reconstruction Level Analysis}
Both the detector effect and beam resolution are considered at the reconstruction level. We produce the truth-level events from Pythia and let them pass through Delphes for a fast detector simulation. The detector configuration is set by the muon collider card in Delphes's package~\cite{Mertens:2015kba,deFavereau:2013fsa,Delphes:muon_card}. The forward muon reconstruction efficiency is approximately 95\%, and the momentum smearing is 10\%. For a 10 TeV muon collider, each beam energy resolution is expected to be $\delta=0.1\%\times 5~$TeV. In order to simulate beam resolution, we generate three different sets of samples with total energy $\{10~\text{TeV}-\sqrt{2}\delta,~10~\text{TeV},~10~\text{TeV}+\sqrt{2}\delta\}$, and combine them with $\{0.25,~0.5,~0.25\}$ weights. Our full reconstruction level analysis is under the following pre-selection cuts before applying our analysis for different detector coverage considerations:
\begin{itemize}
    \item  Only two visible muons in the forward region ($2.5<|\eta(\mu)|<8.0$). 
\item Back-to-back muons: $\eta(\mu^-)\cdot\eta(\mu^+)<0$.
\item Minimal muon momentum: $p_T(\mu)>20~$GeV.
\end{itemize}
Note that depending on the forward coverage of the detector, the minimal momentum $p_T(\mu)$ can be redundant. 

\begin{figure}[htbp]
    \centering
    \includegraphics[width=7.5cm]{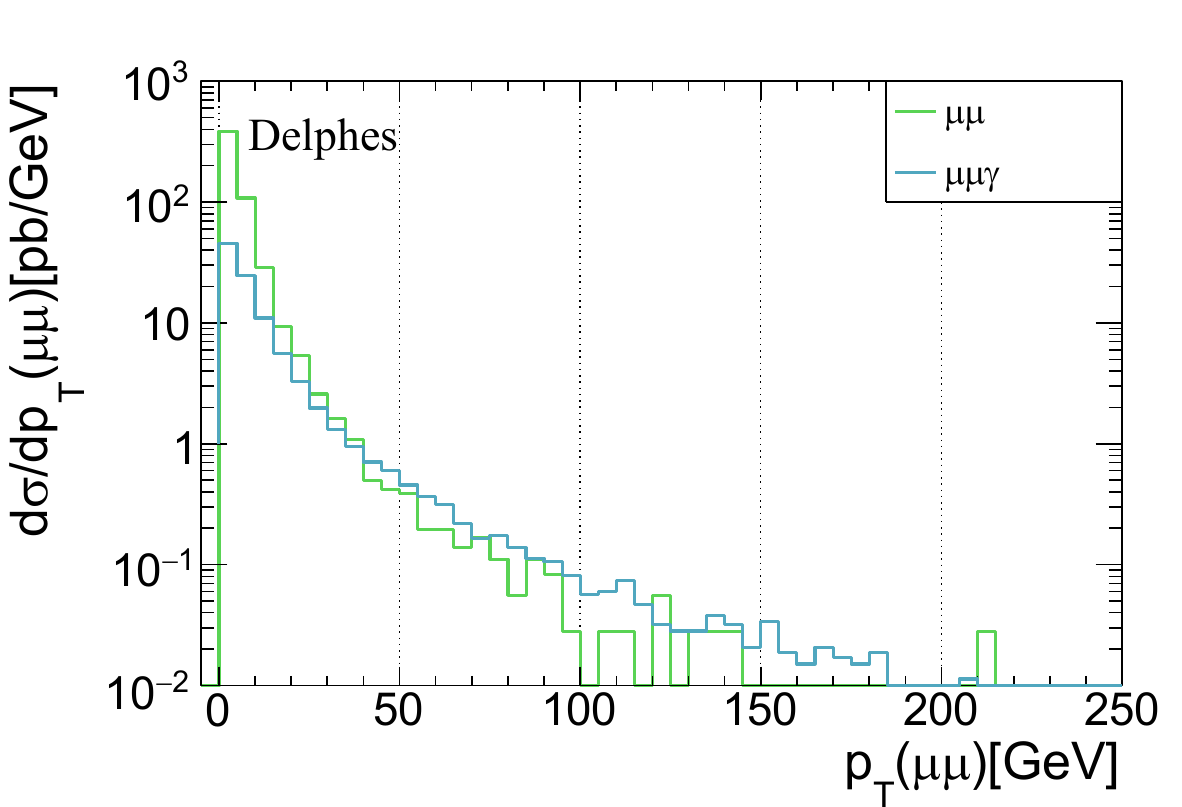}
    \qquad
    \includegraphics[width=7.5cm]{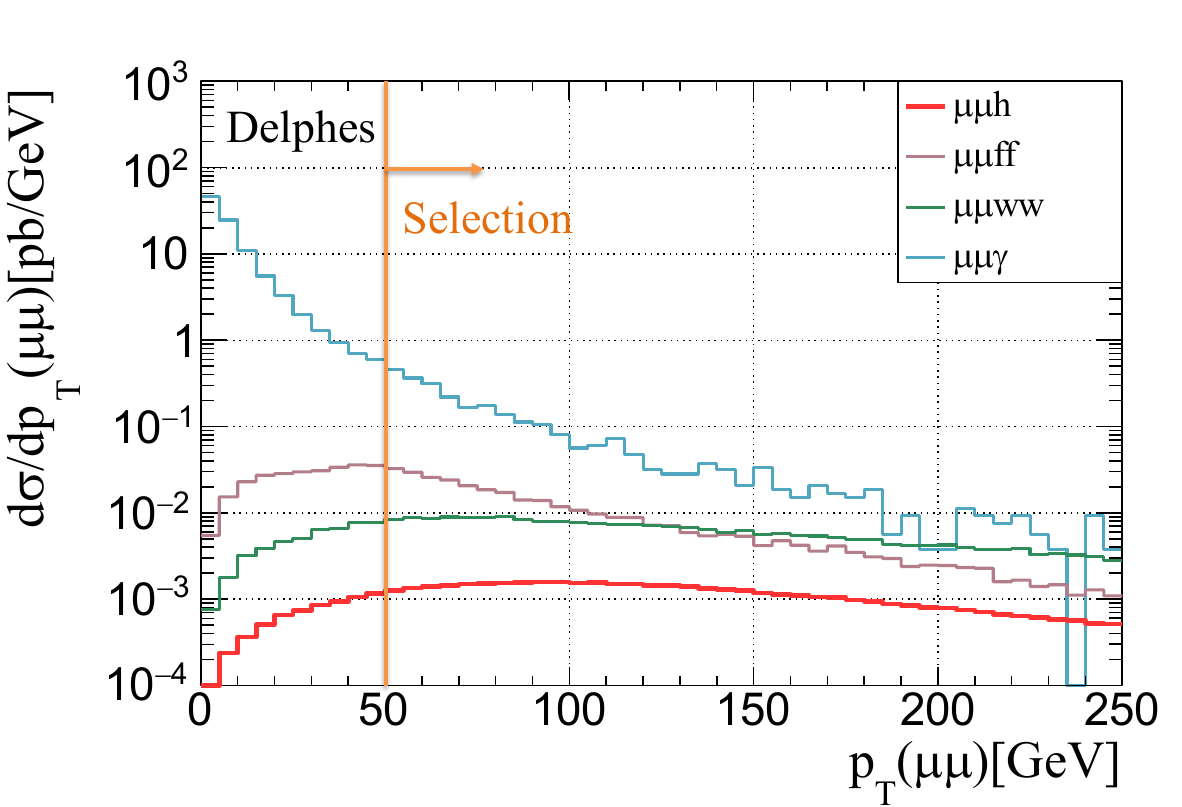}
    \caption{Reconstruction level $p_T(\mu\mu)$ distribution under pre-selection cuts. The left panel compares Bhabha scattering and $\mu^+ \mu^- \to \mu^+ \mu^- \gamma$ process. The right panel shows the $p_T(\mu\mu)$ distribution of all processes.}
    \label{fig:sumPT_mumu_mumua}
\end{figure}

In the previous section, we argue that a cut of $p_T(\mu\mu)>50$ GeV would remove all the pure Bhabha scattering process. Here, we further check the merging pattern between Bhabha scattering and $\mu^+\mu^- \to \mu^+\mu^-\gamma$ under reconstruction level on the right panel of \autoref{fig:sumPT_mumu_mumua}, and observe these two processes crossover at around $35~$GeV, which is still below our choice the transverse momentum cut. Hence, we are further convinced of the safety of ignoring the pure Bhabha scattering after a cut of $p_T(\mu\mu)>50$ GeV. The right plot in \autoref{fig:sumPT_mumu_mumua} shows the $p_T(\mu\mu)$ distribution of all the processes considered.

\begin{table}[ht]
\centering
\resizebox{0.95\textwidth}{!}{%
\begin{tabular}{|c|c|c|c|c|}
\hline
Process & Pre-selection & $p_T(\mu\mu)>50~$GeV & $E(\mu)>3000~$GeV~\&~$p_{T,\text{min}}(\mu)<300~$GeV\\
\hline
$\mu^+ \mu^- \to \mu^+ \mu^- h$ & 73.3\% & 65.7\% & 56.4\%~(0.0489~pb)\\
\hline
$\mu^+ \mu^- \to \mu^+ \mu^- \gamma $ & 13.1\% & 0.38\% & 0.12\%~(0.906~pb)\\
\hline
$\mu^+ \mu^- \to \mu^+ \mu^- f\bar{f} $ & 8.13\% & 4.69\% & 2.58\%~(0.199~pb)\\
\hline
$\mu^+ \mu^- \to \mu^+ \mu^- W^+ W^- $ & 40.0\% & 34.9\% & 22.0\%~(0.207~pb)\\
\hline 
\end{tabular}%
}
\caption{Cutflow table for both signal and background events for 10~TeV muon collider. All processes before the pre-selection cuts are set to 100\%. In the last column, $p_{T,\text{min}}(\mu)$ stands for the transverse momentum of the less energetic forward muon.
}
\label{Table:cutflow_table}
\end{table}

\begin{figure}[htbp]
    \centering 
    \includegraphics[width=0.49\textwidth]{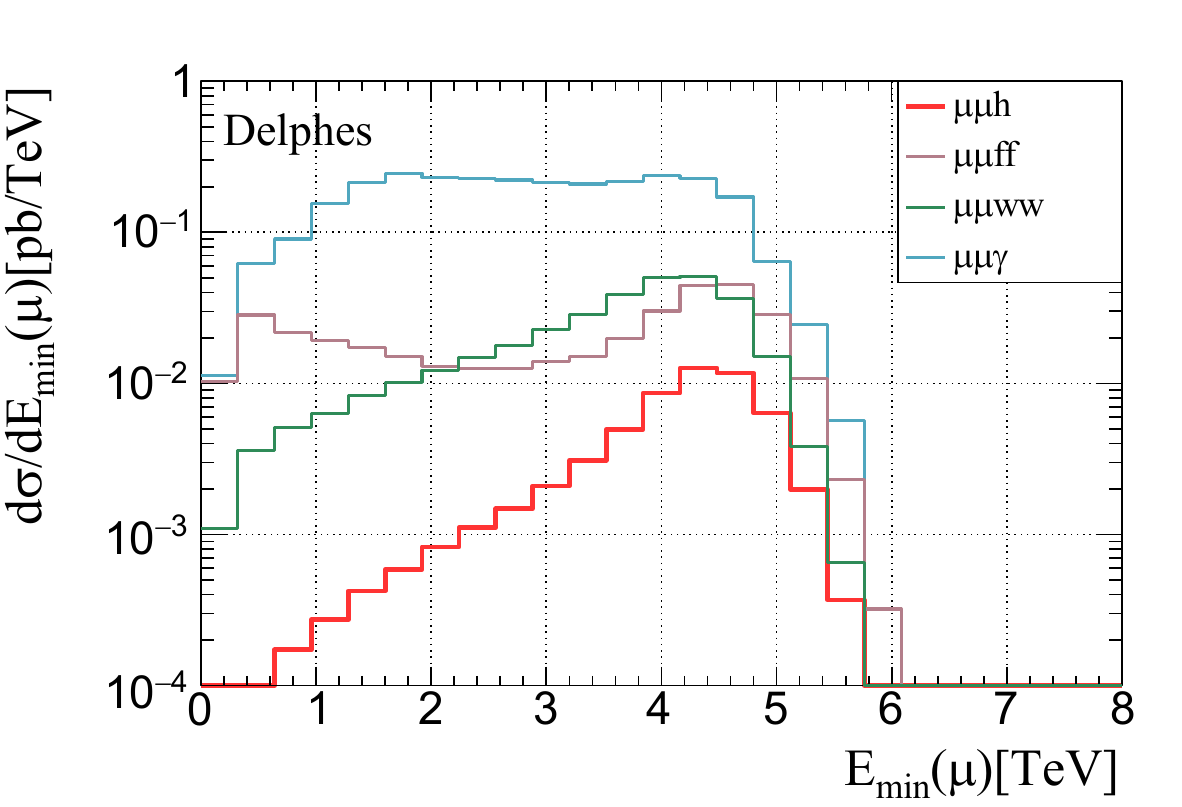}
    \includegraphics[width=0.49\textwidth]{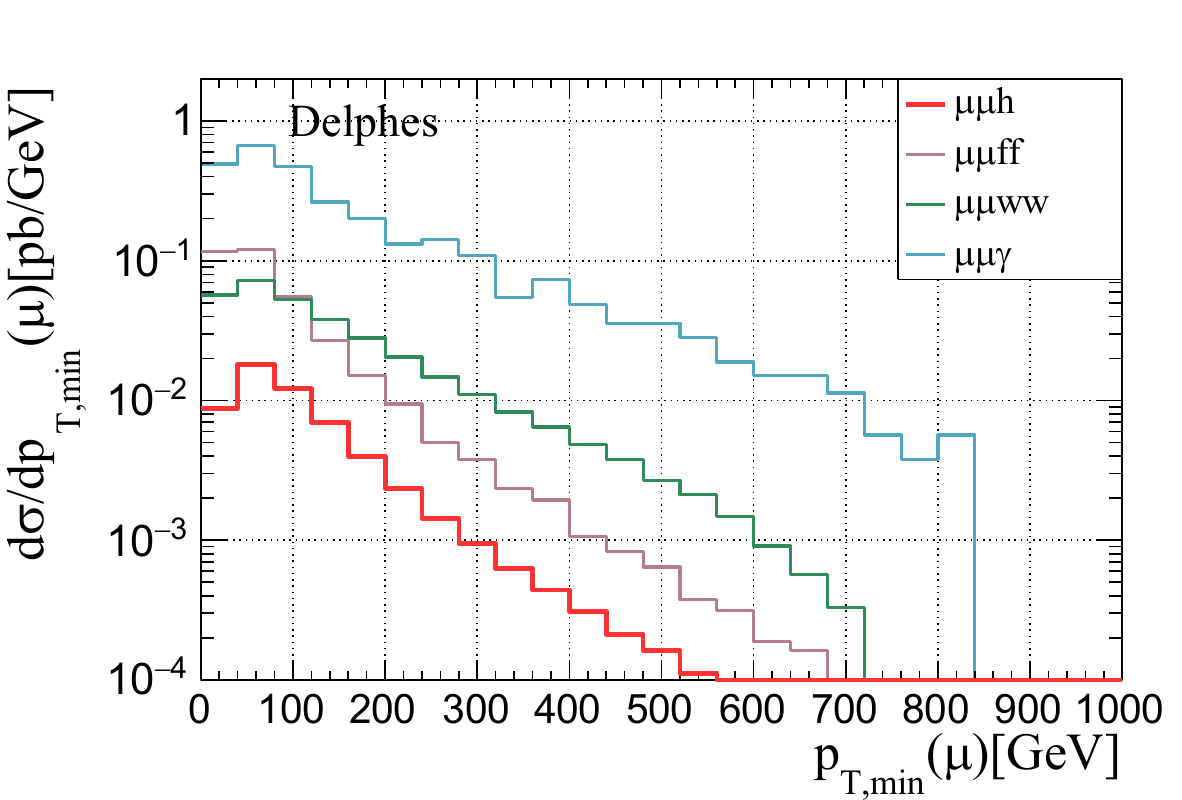}
    \vspace{1ex}
    \includegraphics[width=0.49\textwidth]{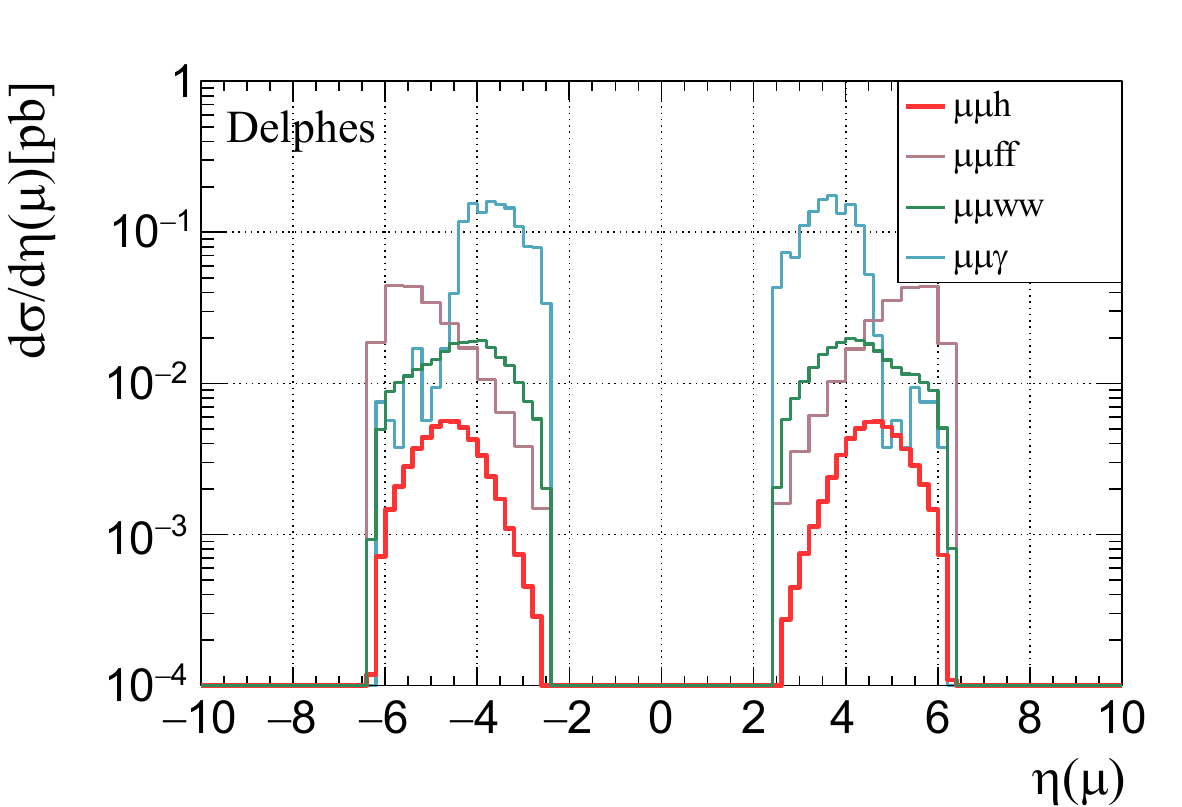}
    \includegraphics[width=0.49\textwidth]{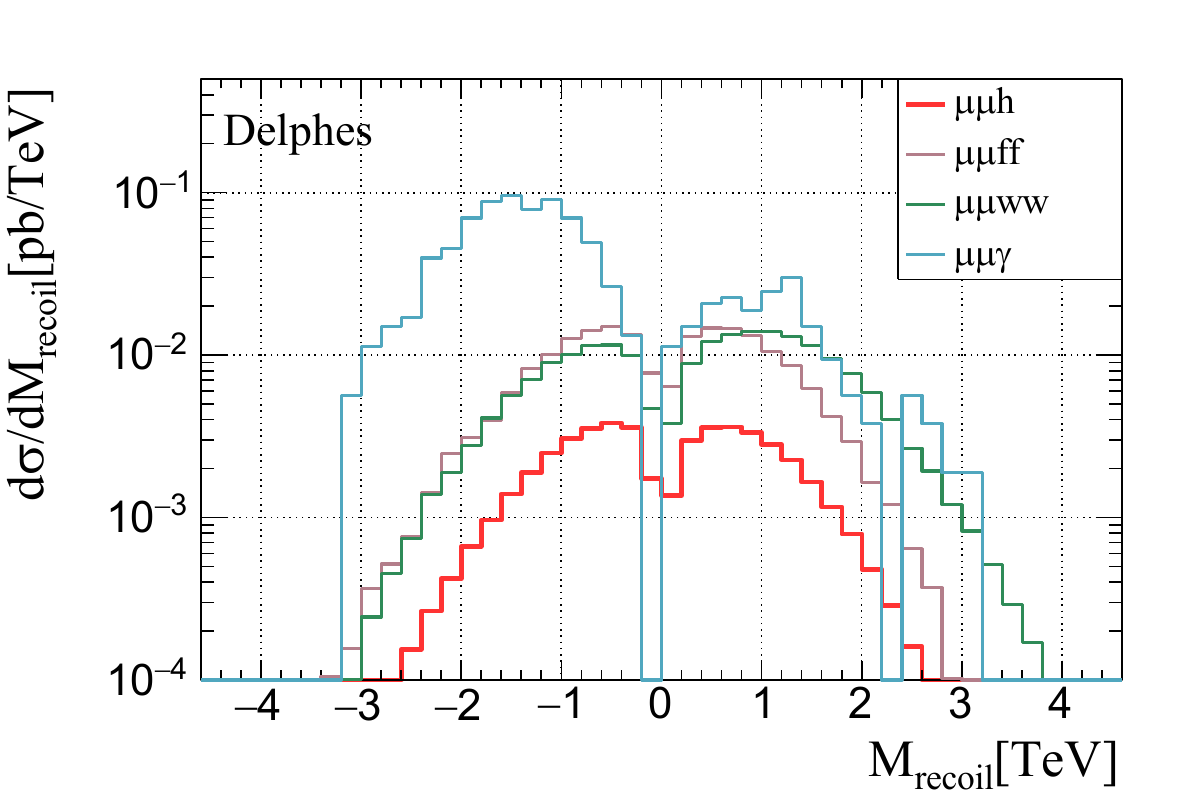}
    \caption{Muon kinematic distributions under reconstruction level. Except for the recoil mass distribution at the right bottom, all other three distributions are after pre-selection and $p_T(\mu\mu)>50~$GeV. The recoil mass distribution is after all the cuts in \autoref{Table:cutflow_table}. The $p_{T,\text{min}}(\mu)$ stands for the transverse momentum of the forward muon, which has lower energy. }
    \label{fig:kinematics_total_bkg}
\end{figure}

After the $p_T(\mu\mu)$ cut, we show several other kinematics distributions in \autoref{fig:kinematics_total_bkg}, and add a cut of $E(\mu)>3000~$GeV and $p_{T,\text{min}}(\mu)<300~$GeV.
A cut-flow table is shown in \autoref{Table:cutflow_table}. Our cut-based analysis keeps around half of the signal, equal to around 0.05 pb in the signal cross-section. While for the three categories of background, only per mil level of $\mu^+ \mu^- \gamma$ is left with a size of around 0.9 pb. For the other two types of background, the remaining cross-sections are each around 0.2 pb. Note that we performed a cut-and-count analysis here for simplicity and robustness of the result, which also leaves room for improvements with multi-variable analysis that leads to better results. For instance, although the recoil mass distribution is wide due to the beam energy spread and reconstruction smearing, we can improve our result by around 5\% by imposing a recoil mass cut. 

\begin{table}[htbp]
\centering
\begin{tabular}{|c|c|c|c|c|}
\hline
Benchmark & $|\eta(\mu)|<4$ & $|\eta(\mu)|<6$ & $|\eta(\mu)|<8$\\
\hline
$\Delta\sigma/\sigma$ & 15\% & 0.75\% & 0.74\% \\
\hline
\end{tabular}%
\caption{The 68\% projected sensitivity on the Higgs inclusive rate from $ZZ$ fusion at 10 TeV muon collider with different forward detector scenarios.}
\label{Table:limit_table}
\end{table}

We carry out the same study for the other two possible forward detector coverage: $|\eta(\mu)|<6$ and $|\eta(\mu)|<4$. 
We perform the study for three possible forward muon detector configurations, which cover $|\eta(\mu)|<8$ (ultra-forward), $|\eta(\mu)|<6$ (forward), and $|\eta(\mu)|<4$ (moderately forward), respectively. 
The precision results are shown in \autoref{Table:limit_table}. For the first narrow range with 
$|\eta(\mu)| < 4$, the $1\sigma$ precision on the $ZZ$ fusion signal rate is just around  $15\%$. While for the other two coverage $|\eta(\mu)| <6$ and $|\eta(\mu)| <8$, the corresponding precision are  $0.75\%$ and $0.74\%$. The result of $|\eta(\mu)|<6$ is much better than $|\eta(\mu)|<4$. The reason can be seen clearly from the $\eta(\mu)$ distribution in \autoref{fig:kinematics_total_bkg}. For a 10~TeV muon collider, most of the signals are accumulated within $|\eta(\mu)|<6$, as one can estimate the signal typical rapidity to be approximately $\log(m_h/4/E_{\rm beam})\simeq 5.1$ or $\log(m_W/2/E_{\rm beam})\simeq 4.8$. This highlights a target forward coverage for a 10~TeV muon collider of $|\eta(\mu)|<6$.

Interestingly, a 3~TeV muon collider would demand less forward coverage. One can perform a similar study and analysis and see that the muons for the signal process are slightly more central. Hence, moderate forward coverage $|\eta(\mu)|<4$ does not significantly differ from the forward muon coverage case of $|\eta(\mu)|<6$. However, the precision would be at around $5\%$ level due to the lower signal rate and luminosity. We show the corresponding results and distributions in Appendix~\autoref{sec:3TeV}.

\section{Higgs Precision}
\label{sec:global_fit}
Our study's primary physics motivation is to highlight that this new, inclusive Higgs channel will enable the high-energy muon collider alone~\footnote{Without relying on any other future Higgs factories, whose timelines are uncertain at this point.}~\footnote{Another muon collider possibility is to have a resonant Higgs factory as a first stage, which can also uniquely determine Higgs width and muon Yukawas to a percent-level precision~\cite{Barger:1996jm,Han:2012rb,deBlas:2022aow}. As shown in Ref.~\cite{Black:2022cth}, before knowing the possibility of the proposed inclusive channel in this paper, the muon collider Higgs fit does not close in a general setup.} to probe Higgs precisely and (by and large) model independently.
\footnote{We also note here that it is possible to use off-shell Higgs rate to constraint the width~\cite{Kauer:2012hd,Forslund:2023reu,Caola:2013yja,Campbell:2015vwa}, which is complementary to our work, but it does depend on the assumption of Higgs coupling scaling at different energy scales~\cite{Logan:2014ppa}.} 

In this section, we perform the global fit to get the projected sensitivity on the Higgs couplings and discuss many new features that emerged from such a new input from this new inclusive channel.

One general and intuitive way to parametrize the precision of the Higgs coupling is the so-called $\kappa$ framework. Compared with the SMEFT framework, this parameterization characterizes the most relevant Higgs coupling deviations while keeping simplicity for describing on-shell Higgs properties. Of course, using the SMEFT would allow one to incorporate the more general comparisons using non-Higgs inputs, such as electroweak precision, and compare Higgs properties across scales~\cite{Buchmuller:1985jz,Grzadkowski:2010es,DiVita:2017vrr,Barklow:2017suo,deBlas:2019rxi,Dawson:2020oco,Ethier:2021bye}. The main point of this study, on the importance of this inclusive channel with forward muon coverage, holds in all Higgs precision frameworks where the Higgs total width is effectively a free parameter~\cite{Dixon:2003yb,Peskin:2012we,Dobrescu:2012td,Kauer:2012hd,Dixon:2013haa,Caola:2013yja,Han:2015ofa,Coradeschi:2015tna,Campbell:2017rke}. Hence, for simplicity, we show our results in the $\kappa$ framework and leave the general SMEFT global fit for future studies. The potential deviation of the SM coupling is expressed as 
\begin{equation}
    \kappa_f = \dfrac{y(h \bar{f} f)}{y_{\rm SM}(h \bar{f} f)}  \, , \quad\quad
    \kappa_V = \dfrac{g(hVV)}{g_{\rm SM}(hVV)}
\end{equation}
with $\kappa =1$ corresponding to the SM prediction. For instance, the major Higgs production channels at high-energy muon colliders are $WW$-fusion and $ZZ$-fusion. One can parameterize the total signal rate under the Narrow Width Approximation (NWA) as
\begin{equation}\label{eq:NWA}
    \sigma(W^* W^* \rightarrow h) \times \text{Br}(h \rightarrow f \bar{f}) = \dfrac{\kappa_W^2 \kappa_f^2}{\kappa_\Gamma} \sigma_{\rm SM}(W^* W^* \rightarrow h) \times \text{Br}_{\rm SM}(h \rightarrow f \bar{f})
\end{equation}
where $\kappa_\Gamma = \Gamma_h/\Gamma_{h,\rm SM}$ being the Higgs total width as a free parameter. For the loop-induced process like $h \rightarrow \gamma \gamma$ that dependents on $\kappa_t$ and $\kappa_W$ from the SM loop computation, we instead choose to parameterize in terms of the effective coupling $\kappa_\gamma$ and $\kappa_{Z\gamma}$, which corresponds to a linear combination of the $H^\dagger H B^{\mu\nu}B_{\mu\nu}$ and $H^\dagger H W^{\mu\nu}W_{\mu\nu}$ operators. In general, new physics considerations and the Higgs portal to the dark sector would induce Higgs invisible decays. Hence, as a typical choice of Higgs global fit, we also include $\Gamma_{\rm inv}$ as a free parameter, which can be parameterized as branching ratio $\text{BR}_{\rm inv}^{\rm BSM}$. Note that there is ``scaling degeneracy'' of the NWA cross-section, which is the origin of why many studies focused on determining the Higgs total width~\cite{Dixon:2003yb,Peskin:2012we,Dobrescu:2012td,Kauer:2012hd,Dixon:2013haa,Caola:2013yja,Han:2015ofa,Coradeschi:2015tna,Campbell:2017rke}. The exclusive Higgs cross-sections \autoref{eq:NWA} are invariant under the scaling of the related coupling by a factor of $t$ and the total decay width scaled by $t^4$. This coupling-width scaling degeneracy for exclusive Higgs observables represents an interesting flat-direction in Higgs global fit, which hinders our extraction of Higgs couplings when projected into space of individual coupling strengths. We emphasize that the overall Higgs property is accessible piece-by-piece in the precision differential cross-section measurements. However, it would appear to be ``lost-in-translation"\footnote{Full information is still kept with the additional correlation matrix.} when expressed in the kappa or SMEFT framework when allowing for general BSM decays that make Higgs width a free parameter. It is unnecessary to overly emphasize this flat direction, as different BSM models would provide additional constraints to enable stronger ones once the underlying model is specified. 
Our study on the inclusive $ZZ$ fusion channel can provide isolated constraints on $\kappa_Z$, thus breaking this flat direction. 

To see the achievable Higgs precision, we take by now a well-established choice in the following $\kappa$ parameters~\footnote{This is close to the kappa-2 and kappa-3 scenarios defined in Ref.~\cite{deBlas:2019rxi}.}
\begin{equation}
\kappa_b,\kappa_t,\kappa_c,\kappa_g,\kappa_W,\kappa_\tau,\kappa_Z,\kappa_\gamma,\kappa_\mu,\text{Br}_\text{inv}^\text{BSM}, \kappa_\Gamma
\end{equation}
with the only requirement $\text{Br}_\text{inv}^\text{BSM} \ge 0$ and all other $\kappa$ parameters can fluctuate around unity. $\kappa_\Gamma$ is a free parameter because one can generally anticipate Higgs has decay channels that are not actively searched for or buried under the background. Generally, any Higgs exotic decays that are not well-constrained or probed can contribute~\cite{Curtin:2013fra,Liu:2016zki,Jung:2021tym}. An equivalent choice is to trade the total width free parameter $\kappa_\Gamma$ by these ``untagged'' branching fractions $\text{Br}_\text{unt}^\text{BSM}$ via
\begin{equation}\label{eq:ktot_relation}
     \kappa_\Gamma = \frac{\sum \kappa_i^2\text{Br}_i^\text{SM}}{1-\text{Br}_\text{inv}^\text{BSM}-\text{Br}_\text{unt}^\text{BSM}}  \  .
\end{equation}
Given that this allows for general, unspecified, new physics contributions, sometimes people call such fitting schemes ``model-independent'' fit. There are no global fits that are not dependent on any assumption, but this scheme does relax the assumptions on Higgs properties in a significant way and covers BSM models broadly. 

There have been many exclusive Higgs sensitivity studies at high-energy muon colliders. We refer to the simulated precision results on various channels adapted in our study as shown in \autoref{Table:sensitivity}, where the effective constrained parameters $\mu$ are defined based on the on-shell Higgs boson production and decay 
\begin{equation}
    \mu_{aa}^{bb}=\frac{\kappa_a^2\kappa_b^2}{\kappa_\Gamma}.
\end{equation}
In this table, the first channel is from Ref.~\cite{Liu:2023yrb} for the indirect probe of the Top Yukawa coupling through the $VV\rightarrow t\bar t$ process. One can also find a compatible, but without all channels and parameterization, study in Ref.~\cite{Chen:2022yiu}. The last channel shown in red is the inclusive channel in this study. The second to last input is from the Higgs invisible study at muon collider in Ref.~\cite{Ruhdorfer:2023uea}. The rest of the input precisions on Higgs exclusive decay channels are collected from Ref~\cite{Forslund:2022xjq}. 
We show the constraint results for the 11 free parameters under both $\kappa$ frameworks at 10 TeV muon collider in \autoref{Table:sensitivity_upper_lower}. 
We verified that the results are the same for all the fitting parameters at two significant digit levels for these two fitting choices. 
Hence, we incorporate all the results into one column. The two branching ratio values are presented to be the exclusion rate at 95\% C.L. We mainly focus on the forward muon result at $\eta(\mu)<6$ as this provides almost optimal performance, and in the following discussions about the results, we focus on this scenario. We also provide the global fit result with $\eta(\mu)<4$ for comparison, and as we can see, the result is generally about one order of magnitude worse. Hence, we should design a forward muon detection with coverage up to  $\eta(\mu)$ of 6 for Higgs physics purposes.

\begin{table}[t]
\centering
\begin{tabular}{|c|c|c|c|c|c|c|c|c|c|c|c|}
\hline
$\mu_{\text{production}}^{\text{decay}}$ & $\mu_{VV}^{tt}$ & $\mu_{WW}^{bb}$ & $\mu_{WW}^{cc}$ & $\mu_{WW}^{gg}$ & $\mu_{WW}^{\tau\tau}$ & $\mu_{WW}^{WW}$ & $\mu_{WW}^{ZZ}$ & $\mu_{WW}^{\gamma\gamma}$ & $\mu_{WW}^{\mu\mu}$\\
\hline
$\Delta\sigma/\sigma (\%)$ & $2.8$
& $~0.22~$ & $~3.6~$ & $~0.79~$ & $~1.1~$ & $~0.40~$ & $~3.2~$ & $~1.7~$ & $~5.7~$ \\
\hline 
$\mu_{\text{production}}^{\text{decay}}$ & $\mu_{ZZ}^{bb}$ & $\mu_{ZZ}^{cc}$ & $\mu_{ZZ}^{gg}$ & $\mu_{ZZ}^{\tau\tau}$ & $\mu_{ZZ}^{WW}$ & $\mu_{ZZ}^{ZZ}$ & $\mu_{ZZ}^{\gamma\gamma}$ & $\mu_{ZZ}^\text{inv}$ &  \color{red}{$\mu_{ZZ}^{H}$}\\
\hline
$\Delta\sigma/\sigma (\%)$ & $~0.77~$ & $~17~$ & $~3.3~$ & $~4.8~$ & $~1.8~$ & $~11~$ & $~4.8~$ & $~0.05~$ &  \color{red}{~0.75~} \\
\hline 
\end{tabular}
\caption{The input precision used in this study for the Higgs precision fit for a 10~TeV High Energy Muon Collider at 10~ab$^{-1}$, and the $ZZ$-fusion channel assumes forward coverage. The inclusive Higgs channel studied here is shown in red. The cross-section precision including Top Yukawa coupling determination~\cite{Liu:2023yrb}, various Higgs decay channel~\cite{Forslund:2022xjq}, Higgs invisible decay study~\cite{Ruhdorfer:2023uea} and our study of inclusive Higgs rate at 10 TeV Muon Collider. 
}
\label{Table:sensitivity}
\end{table}

\begin{table}[htbp]
\centering
\begin{tabular}{|l|ccc|ccc|}
\hline
                           &    \multicolumn{3}{c|}{$|\eta(\mu)|<4$}& \multicolumn{3}{c|}{$|\eta(\mu)|<6$}\\
                           & MuC@10TeV& +HL-LHC& ~+$e^+e^-$~& MuC@10TeV& +HL-LHC& ~+$e^+e^-$~\\ \hline
\multirow{2}{*}{$\kappa_b(\%)$} &    $+7.5$&    $+1.7$&    $+0.25$& $+0.56$&         $+0.53$&          $+0.24$\\
                           &    $-0.25$&    $-0.24$&    $-0.18$& $-0.23$&         $-0.23$&          $-0.17$\\ \hline
\multirow{2}{*}{$\kappa_t(\%)$} &    $+1.4$&    $+1.3$&    $+1.3$
& $+1.4$&         $+1.3$&          $+1.3$\\
                           &    $-7.1$&    $-1.6$&    $-1.2$& $-1.4$&         $-1.2$&          $-1.2$\\ \hline
\multirow{2}{*}{$\kappa_c(\%)$}&    $+7.8$&    $+2.6$&    $+0.91$& $+1.8$&         $+1.8$&          $+0.89$
\\
                           &    $-2.1$&    $-2.1$&    $-0.91$& $-1.8$&         $-1.8$&          $-0.89$
\\ \hline
\multirow{2}{*}{$\kappa_g(\%)$}&    $+7.5$&    $+1.7$&    $+0.38$& $+0.67$&         $+0.63$&          $+0.35$
\\
                           &    $-0.52$&    $-0.50$&    $-0.35$& $-0.45$&         $-0.44$&          $-0.32$
\\ \hline
\multirow{2}{*}{$\kappa_W(\%)$}&    $+7.5$&    $+1.7$&    $+0.17$& $+0.51$&         $+0.48$&          $+0.16$
\\
                           &    $-0.15$&    $-0.13$&    
$-0.099$& $-0.10$&         $-0.10$&          $-0.090$
\\ \hline
\multirow{2}{*}{$\kappa_\tau(\%)$}&    $+7.5$&    $+1.8$&    $+0.33$& $+0.76$&         $+0.71$&          $+0.32$
\\
                           &    $-0.62$&    $-0.57$&    $-0.27$& $-0.56$&         $-0.55$&          $-0.27$
\\ \hline
\multirow{2}{*}{$\kappa_Z(\%)$}&    $+7.3$&    $+1.9$&    $+0.13$& $+0.37$&         $+0.37$&          $+0.12$
\\
                           &    $-1.4$&    $-0.93$&    $-0.058$& $-0.25$&         $-0.25$&         $ -0.056$
\\ \hline
\multirow{2}{*}{$\kappa_\gamma(\%)$}&    $+7.6$&    $+1.8$&    $+0.66$& $+0.97$&         $+0.86$&          $+0.65$
\\
                           &    $-0.83$&    $-0.71$&    $-0.64$& $-0.82$&         $-0.71$&          $-0.64$\\ \hline
\multirow{2}{*}{$\kappa_\mu(\%)$}&    $+9.1$&    $+3.8$&    $+2.3$& $+2.9$&         $+2.5$&          $+1.9$\\
                           &    $-5.0$&    $-3.6$&    $-2.4$& $-2.9$&         $-2.5$&          $-2.0$\\ \hline
\multirow{2}{*}{$\text{Br}_\text{inv}^{95\%}(\%)$}&    $+0.64$&    $+0.63$&    $+0.13$& $+0.10$&         $+0.10$&          $+0.080$\\
                           &    0
&    0
&    0
& 0&         0&          0\\ \hline
\multirow{2}{*}{$\text{Br}_\text{unt}^{95\%}(\%)$}&    $+27$&    
$+6.6$&    $+0.57$& $+2.0$&         $+1.9$&          $+0.54$\\
                           &    0&    0&    
0& 0&         0&          0\\ \hline \hline
\multirow{2}{*}{$\kappa_\Gamma(\%)$}&    $+34$&    
$+6.9$&    $+0.69$& $+2.1$&         $+1.9$&          $+0.65$\\
                           &    $-0.45$&    $-0.43$&    $-0.31$& $-0.41$&         $ -0.40$&          $-0.29$\\ \hline
\end{tabular}%
\caption{The $1\sigma$ upper and lower bound of the 11 parameters in the Higgs global fit under two different forward coverage at 10 TeV MuC. The constraint values for $\text{Br}_\text{inv}$ and $\text{Br}_\text{unt}$ are the exclusion rate at 95\% C.L.
}
\label{Table:sensitivity_upper_lower}
\end{table}

Except for the two branching ratios, which can be only larger or equal to zero, all the other $\kappa$ parameters can be on either side of unity. In other words, $\Delta \kappa$ can take either positive or negative values. However, the constraint results in \autoref{Table:sensitivity_upper_lower} for these $\Delta \kappa$ are not symmetric around zero due to the asymmetric boundary conditions. The negative branch can achieve better precision due to scaling degeneracy and physical boundary conditions.
As we stated before, the observable $\mu$ is invariant under the scaling $\kappa_f \rightarrow t\kappa_f,\kappa_V \rightarrow t\kappa_V$ as well as $\kappa_\Gamma \rightarrow t^4\kappa_\Gamma$.  As shown in \autoref{eq:ktot_relation}, $\kappa_\Gamma$ scales like $\kappa_{f/V}^2$. If $\kappa_{f/V} > 1$, the degeneracy condition requires $\kappa_\Gamma \sim \kappa_{f/V}^4$, which can be compensated by positive $\text{Br}_\text{unt}$ value. On the contrary, if $\kappa_{f/V} < 1$, $\kappa_\Gamma$ is smaller than the quadratic scaling, which requires negative $\text{Br}_\text{unt}$ value. However, physically, the branching ratio can only be equal to or larger than zero. 
Hence, the degeneracy condition cannot be satisfied for the $\kappa$ parameters shown in the highly precise observables. As a result, on the negative branch, the 68\% allowed $\kappa$ parameters must be determined by the relative error of the $\mu$ observables. This generally enhances the corresponding projected sensitivity for the $\kappa_{f/V}$ compared to the positive branch, except the $\kappa_t$ under $|\eta(\mu)|<4$ scenario. From the Top Yukawa coupling determination~\cite{Liu:2023yrb}, the $\Delta\chi^2$ only depends on 
\begin{equation}
    O(10^3)\times(\kappa_{V}\kappa_t-1)^n\approx O(10^3)\times(\Delta\kappa_{V}+\Delta\kappa_t)^n.
\end{equation}
The sum of $\Delta\kappa_{V}$ and $\Delta\kappa_t$ need to be small, which implies that once the deviation of $\Delta\kappa_V$ is large, it will dominantly determine the opposite deviation of $\Delta\kappa_t$. It is the case that, under $\eta<4$ scenario, both $\Delta\kappa_Z$ and $\Delta\kappa_W$ can positively deviate to $\sim7\%$ due to the poor precision shown in \autoref{Table:limit_table}.

Here, we provide some approximation to help the reader understand the result, comment on special features unique to muon colliders, and avoid viewing the Higgs global fit result as a black box.
Let us first focus on the positive branch of the global fit. The 68\% precision of $\Delta\kappa_Z$ equals half of $\mu_{ZZ}^H$, as this is the key and leading constraints to confine the overall scale of the Higgs couplings, exactly as what we anticipated and motivated this study. The $\kappa_\Gamma$ can be constrained to be around 2\%. We can get a reasonable estimation on $\kappa_\Gamma$ via picking up four observables dependent on $\kappa_b,\kappa_W$,$\kappa_Z$ and $\kappa_\Gamma$. Their correlation gives
\begin{equation}
    \kappa_\Gamma = \dfrac{\brr{\mu_{ZZ}^H}^2}{ \mu_{WW}^{WW} } \brr{ \dfrac{\mu_{WW}^{bb} }{\mu^{bb}_{ZZ} } }^2
\end{equation}
which gives
\begin{equation}
    \Delta \kappa_\Gamma = \srr{ 4 \brr{\Delta \mu_{ZZ}^H}^2 + \brr{\Delta \mu_{WW}^{WW}}^2 + 4(\Delta\mu_{WW}^{bb})^2+4(\Delta\mu_{ZZ}^{bb})}^{1/2} = 2.2 \%
\end{equation}
Incorporating other observables and parameters would further improve the sensitivity of $\kappa_\Gamma$. On the other hand, the size of $\Delta\kappa_W$ and $\Delta\kappa_b$ are to be one-quarter of $\Delta \kappa_\Gamma$. 
For instance, the leading constraints on $\kappa_W$ can be derived from
\begin{equation}
    \kappa_W^4=(\mu_{WW}^{WW})\kappa_\Gamma={\brr{\mu_{ZZ}^H}^2} \brr{ \dfrac{\mu_{WW}^{bb} }{\mu^{bb}_{ZZ} } }^2,
    \label{eq:kappaW}
\end{equation}
implying
\begin{equation}
    \Delta \kappa_W = \frac 1 4 \srr{ 4 \brr{\Delta \mu_{ZZ}^H}^2 + 4(\Delta\mu_{WW}^{bb})^2+4(\Delta\mu_{ZZ}^{bb})^2}^{1/2} = 0.55 \%.
\end{equation}
Similarly, the leading constraints on $\kappa_b$ can be derived from
\begin{equation}
    \kappa_b^2=\frac{\mu_{WW}^{bb}\kappa_W^2} {\mu_{WW}^{WW}}= \frac{\mu_{ZZ}^H (\mu_{WW}^{bb} )^2} {\mu^{bb}_{ZZ} \mu_{WW}^{WW}},
\end{equation}
implying
\begin{equation}
    \Delta \kappa_b = \frac 1 2 \srr{ \brr{\Delta \mu_{ZZ}^H}^2 + 4(\Delta\mu_{WW}^{bb})^2+(\Delta\mu_{ZZ}^{bb})^2+(\Delta\mu_{WW}^{WW})}^{1/2} = 0.61 \%.
\end{equation}
This is due to the two most precise observables $\mu_{WW}^{bb}$ and $\mu_{WW}^{WW}$. 
The constraint for $\kappa_\mu,\kappa_c,\kappa_\gamma$ is roughly to half of the precision of $\mu_{WW}^{\mu\mu},\mu_{WW}^{cc}$ and $\mu_{WW}^{\gamma\gamma}$ as the corresponding exclusive cross-section dominant the uncertainty. In other channels, such as $\kappa_\tau$ and $\kappa_g$, one can use a similar derivation as $\kappa_b$ above to see the leading contribution, where uncertainties of $\kappa_W$ and $\mu_{WW}^{WW}$ contributes. 
The $\kappa_t$ is from an interference determination, where $\kappa_W$ and $\kappa_Z$ also enter. 

The results from the above reveal an interesting feature: the extracted coupling precision for Higgs is generally better than half of the total width precision. This sharply contrasts the results from many future Lepton collider Higgs factories, such as CEPC, FCC-ee, ILC, and C3. For these colliders, the main channel to break the scaling degeneracy is the inclusive $Z$ of the Higgs-associated production, which is also the major production channel of the Higgs, but the process of $\mu_{ZH}^{ZZ}$ cannot be determined to be much better than $\kappa_\Gamma$. The resulting coupling precision is generally larger than half of $\kappa_\Gamma$, through $\kappa_{x}=\kappa_\Gamma \mu_{ZH}^{xx} /\mu_{ZH}$. For high-energy muon colliders, the major production coupling $\kappa_W$ can be determined via the chain in \autoref{eq:kappaW}. One can then use cross-section ratio with $\mu_{WW}^{WW}$ and $\kappa_W^2$ to get rid of direct dependence on $\kappa_\Gamma$, which yields a better precision than one would naively anticipate. One can summarize the situation: the high precision in different production modes at high-energy muon colliders enables a better Higgs precision through diversification. One also observes similar effects and realizes the importance of the $WW$-fusion run at the 360~GeV electron-positron colliders. 

The constraint results differ when switching to the negative branch of the Higgs precision. The scaling degeneracy is broken due to the non-negative value of BSM decay partial widths. The boundary of the Higgs precision is mostly set by minimal total width case, where all BSM decay particle widths are turned off. Similar to the positive branch discussion, while the global fit is performed in the fully general setup, we can have some analytical insights to understand the results. The first set of constraints is mainly from $\mu_{WW}^{bb}$ and $\mu_{WW}^{WW}$. The lower boundary on $\kappa_W$ is set to be within uncertainty of $\mu_{WW}^{WW}$, which would simultaneously require the minimization of $\kappa_b$, as the Higgs branching to $b\bar b$ in SM is 57.8\%, and to $WW$ is 21.6\%, dominating the Higgs width. The lower boundary on $\kappa_W$ also needs to be within the uncertainty of $\mu_{WW}^{bb}$, which would require maximizing $\kappa_b$. Noticing that, 
\begin{equation}
\Br_b^{\rm SM} \mu_{WW}^{bb}+\Br_W^{\rm SM} \mu_{WW}^{WW}= \kappa_W^2 \frac {\kappa_b^2 \Br_b^{\rm SM} +\kappa_W^2 \Br_W^{\rm SM}} {\kappa_b^2 \Br_b^{\rm SM} +\kappa_W^2 \Br_W^{\rm SM} + ...}\simeq \kappa_W^2,
\end{equation}
if one makes the approximation to ignore (and the variations) all other SM partial widths, one can find, 
$\Delta\kappa_W\simeq -0.11\%$. Note that the sum of the rest of SM branching would not vary much as each coupling would vary less than the \% level (as one can derive from the general equations when discussing the results in the positive branch), so this is a somewhat reasonable approximation. We note here that the discussion on the negative branch of results is less rigorous as its boundary condition is more complex and involves all Higgs couplings. On the other hand, such an approximation is no longer valid when discussing the positive side of Higgs precision.
$\kappa_b$ can be approximately understood by using $\mu_{WW}^{WW}$ and the lower bound on $\kappa_W$ derived earlier. One can also obtain the results by performing a two-parameter, $\kappa_W$ and $\kappa_b$ with input precision of $\mu_{WW}^{WW}$ and $\mu_{WW}^{bb}$, assuming all other width variations can be ignored to obtain similar results above.

\begin{figure}[t]
    \centering
    \includegraphics[width=16cm]{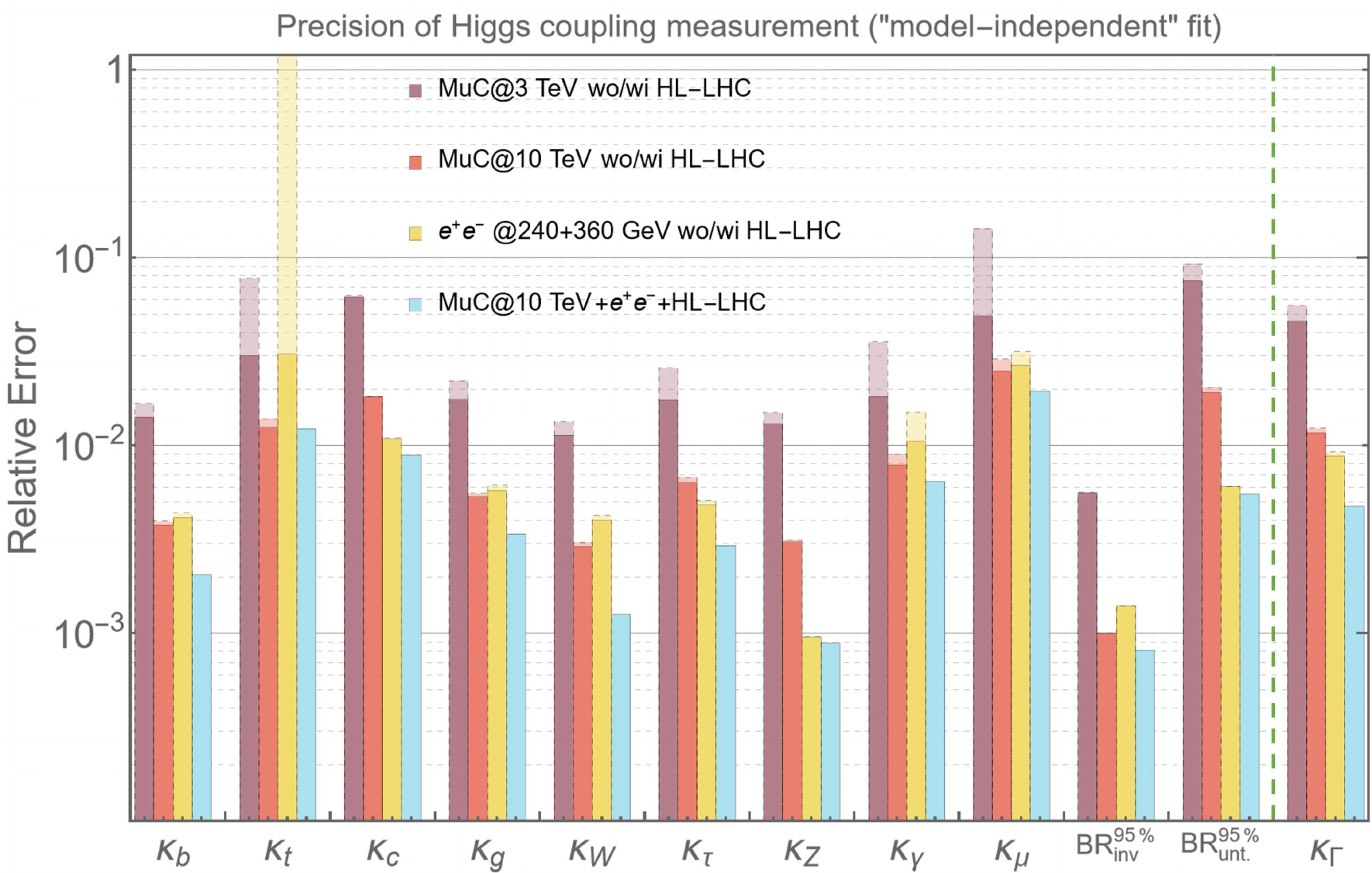}
    \caption{
    11-parameter Higgs global fit where the constraint on total width $\kappa_\Gamma$ is determined from the previous 11 parameters. The muon collider results are under the forward detection $|\eta(\mu)|<6$.
    The light-colored bars stand for the results without combining with HL-LHC. The global fit at CEPC without HL-LHC is under a 10-parameter fit (excluding $\kappa_t$) to avoid a flat direction.
    }
    \label{fig:parameter_fit}
\end{figure}

In \autoref{fig:parameter_fit}, we show the results for different high energy muon collider scenarios, $e^+e^-$ Higgs factories, and combined with HL-LHC\footnote{The HL-LHC precision results (Scenario 2) are taken from  \href{https://twiki.cern.ch/twiki/bin/view/LHCPhysics/GuidelinesCouplingProjections2018}{https://twiki.cern.ch/twiki/bin/view/LHCPhysics/GuidelinesCouplingProjections2018}.} and $e^+e^-$ Higgs factories\footnote{Here we use the CEPC input precision~\cite{CEPCStudyGroup:2018ghi,An:2018dwb} as a representative precision for $e^+e^-$ Higgs factories, where in details ILC, C3, FCC-ee and CEPC differ slightly.}.
To avoid the complexities of asymmetric uncertainties and as many inputs for other colliders are missing, we show the symmetrized uncertainty here, averaging the absolute precision value in the positive and negative directions. The last bar displays the constraint for the $\kappa_\Gamma$, highlighting its signature and importance. One can extract many interesting messages from this plot.
With our proposed channel and forward coverage, one can obtain a closed fit at a 3~TeV muon collider alone. The general achievable precision is at \% level, a factor of a few worse than a typical $e^+e^-$ Higgs factory. On the other hand, a 10~TeV muon collider alone would achieve sub-percent level precision for many couplings, comparable or, in some cases, better than $e^+e^-$ Higgs factories. We found that combining with the HL-LHC would not improve the Higgs precision much for a 10~TeV muon collider or $ e^+e^- $ Higgs factories, which is different for the 3 TeV muon collider. The feature of $e^+e^-$ Higgs factories lies in the measurement of the Higgsstrahlung process, which can limit $\kappa_Z$ as low as around 0.1\%. Due to the center-of-mass energy limit, $e^+e^-$ Higgs factories (except for CLIC or C3) cannot constrain $\kappa_t$, manifested by the light yellow bar. After combining with HL-LHC simulated observable, $\kappa_t$ can be measured to be around 3\%, similar to 3 TeV muon collider. While the 10 TeV muon collider can probe $\Delta\kappa_t$ to around 1.3\%, benefiting from the indirect channels and higher luminosity. 
Notably, although the Higgs total width precision at a 10~TeV muon collider, enabled by our proposed channel with forward muon coverage, is worse than that of $e^+e^-$ Higgs factories, the resulting coupling precision is in many cases better, in particular $\kappa_b$ and $\kappa_W$. Lastly, as shown in the cyan bars, combining high energy muon collider and $e^+e^-$ machine would improve many precisions significantly, more than naive statistical additions. This demonstrates the complementarity between different colliders and shows us the importance of producing and detecting Higgs in many channels. 
Our proposed inclusive measurement relies on a non-trivial development of a forward muon detector but delivers a robust Higgs width and coupling determination on-shell. Note that one can also determine the Higgs overall coupling through the on-shell, off-shell Higgs measurements~\cite{Kauer:2012hd,Forslund:2023reu,Caola:2013yja,Campbell:2015vwa} at muon colliders~\cite{Forslund:2023reu}. 
Such a determination of Higgs width and coupling strength benefits from not relying on forward muon tagging. However, it depends on the assumption of Higgs' boson coupling modification being invariant across the scales of the measurements~\cite{Logan:2014ppa}. In the same fitting framework, our determination of Higgs couplings in many cases is a factor of few better than this complementary measurement (see Fig.7 of Ref.~\cite{Forslund:2023reu}).

\section{Conclusion}
\label{sec:conclusion}

We aim to solve the shortcoming of high energy muon collider, which is anticipated not to be able to break the Higgs coupling-width degeneracy or often simplified as incapable of determining the Higgs boson width. We show that through the $ZZ$-fusion channel with a forward muon detection, one can measure the Higgs rate {\it inclusively}. Such a measurement breaks the coupling-width degeneracy and enables the determination of the Higgs coupling to sub-percent level and the total width to $(-0.41\%,+2.1\%)$ at 10~TeV muon collider alone.
The results are summarized in \autoref{Table:sensitivity_upper_lower} and \autoref{fig:parameter_fit}. Here, we highlight a few key findings. Due to the multiple Higgs production modes and low background, a high-energy muon collider is also highly complementary to other Higgs factories, combining them to improve the Higgs precision significantly. Note that we performed a cut-and-count analysis here, which can be improved with multi-variable analysis that leads to better results. 

Forward muon tagging is critical for such inclusive Higgs measurements. 
While one naively considers the recoil mass to provide a sharp peak to separate signal and background, the muon energy reconstruction and beam energy spread render such distribution rather broad, invalidating the potential strong usage of such a variable. However, being a lepton collider, the muon collider is still clean and enjoys a low background. We found that a signal-background ratio of around 1/30 is achievable. The critical implication is that one should be able to tag the high-energy muons without high demand on the accuracy of energy reconstruction and the energy loss in the shielding region.

For a 10~TeV muon collider, most forward muons reside with $|\eta(\mu)|<6$. Covering this region improves the precision by a factor of 20, as shown in \autoref{Table:limit_table}. For a 3~TeV muon collider, the muons are slightly less forward but still benefit from having a forward muon detector, as shown in \autoref{Table:limit_3TeV_table}. 

The inclusive Higgs measurement is unique for high-energy lepton colliders; it enables often-dubbed ``model-independent'' Higgs coupling determination where the total width is a free parameter. New studies, including forward muon detector design and performance estimation, precision timing, beam-induced backgrounds, and precision calculation of the signal and background rates, are all highly motivated future steps to fully realize the high energy muon collider's Higgs physics potential.

\acknowledgements{We thank Matthew Forslund, Simone Pagan Griso, Tao Han, Donatella Lucchesi, Patrick Meade, and Andrea Wulzer for helpful discussions. This study was supported in part by the DOE Grant DE-SC0022345 and DE-SC0011842. Z.L. and K.F.L acknowledge KITP for hosting the muon collider workshop, where this study was inspired, which was supported in part by the National Science Foundation under Grant No. NSF PHY-1748958. Z.L. and K.F.L. acknowledge the
support of the Aspen Center for Physics, supported by National Science Foundation grant PHY-2210452, where part of this work was completed. The data associated with the figures in this paper can be accessed via \href{https://github.com/ZhenLiuPhys/MuCHiggsInclusive}{Github~\faGithub}. }

\appendix

\section{$ZZ$-fusion for the Inclusive Rate at 3 TeV Muon Collider}
\label{sec:3TeV}

\begin{figure}[htbp]
    \centering
    \includegraphics[width=9.0cm]{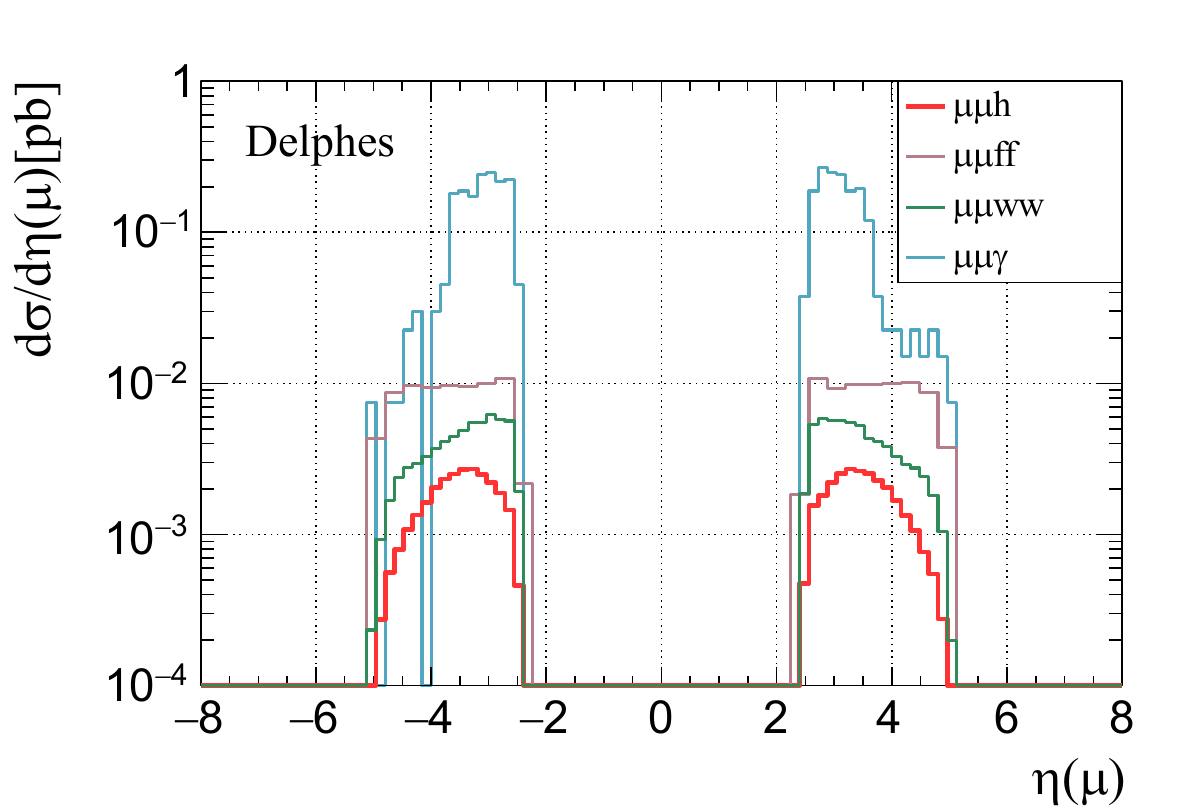}
    \caption{The reconstruction level $\eta(\mu)$ distribution are after pre-selection and $p_T(\mu\mu)>50~$GeV at 3 TeV muon collider.}
    \label{fig:ForwardMuon_Eta_3_TeV}
\end{figure}

\begin{table}[htbp]
\centering
\resizebox{0.5\textwidth}{!}{%
\begin{tabular}{|c|c|c|c|c|}
\hline
Benchmark & $|\eta(\mu)|<4$ & $|\eta(\mu)|<6$ & $|\eta(\mu)|<8$\\
\hline
$\Delta\sigma/\sigma$ & 6.2\% & 3.9\% & 3.9\% \\
\hline
\end{tabular}%
}
\caption{The 68\% projected sensitivity on the Higgs inclusive rate from $ZZ$ fusion at 3 TeV Muon Collider with different forward detector scenarios.}
\label{Table:limit_3TeV_table}
\end{table}

\begin{table}[htbp]
\centering
\begin{tabular}{|l|ccc|ccc|}
\hline
                           &    \multicolumn{3}{c|}{$|\eta(\mu)|<4$}& \multicolumn{3}{c|}{$|\eta(\mu)|<6$}\\
                           & MuC@3TeV& +HL-LHC& ~+$e^+e^-$~& MuC@3TeV& +HL-LHC& ~+$e^+e^-$~\\ \hline
\multirow{2}{*}{$\kappa_b(\%)$} &    $+6.5$&    $+2.7$&    $+0.37$& $+2.5$&         $+2.0$&          $+0.36$\\
                           &    $-0.90$&    $-0.81$&    $-0.30$& $-0.86$&         $-0.79$&          $-0.29$\\ \hline
\multirow{2}{*}{$\kappa_t(\%)$} &    $+9.1$&    $+3.5$&    $+2.8$& $+9.2$&         $+3.2$&          $+2.8$\\
                           &    $-8.6$&    $-2.8$&    $-2.8$& $-6.4$&         $-2.8$&          $-2.8$\\ \hline
\multirow{2}{*}{$\kappa_c(\%)$}&    $+9.2$&    $+6.9$&    $+1.0$& $+6.3$&         $+6.1$&          $+1.0$\\
                           &    $-6.9$&    $-6.9$&    $-1.0$& $-6.3$&         $-6.3$&          $-1.0$\\ \hline
\multirow{2}{*}{$\kappa_g(\%)$}&    $+6.7$&    $+3.0$&    $+0.52$& $+2.8$&         $+2.3$&          $+0.51$\\
                           &    $-1.9$&    $-1.4$&    $-0.49$& $-1.6$&         $-1.3$&          $-0.48$\\ \hline
\multirow{2}{*}{$\kappa_W(\%)$}&    $+6.6$&    $+2.7$&    $+0.30$& $+2.3$&         $+1.9$&          $+0.29$\\
                           &    $-0.46$&    $-0.39$&    
$-0.26$& $-0.37$&         $-0.35$&          $-0.25$\\ \hline
\multirow{2}{*}{$\kappa_\tau(\%)$}&    $+6.8$&    $+2.9$&    $+0.44$& $+3.1$&         $+2.2$&          $+0.44$\\
                           &    $-2.3$&    $-1.3$&    $-0.38$& $-2.1$&         $-1.3$&          $-0.38$\\ \hline
\multirow{2}{*}{$\kappa_Z(\%)$}&    $+3.1$&    $+2.5$&    $+0.13$& $+2.0$&         $+1.8$&          $+0.13$\\
                           &    $-2.7$&    $-1.2$&    $-0.063$& $-1.0$&         $-0.82$&         $ -0.063$\\ \hline
\multirow{2}{*}{$\kappa_\gamma(\%)$}&    $+7.2$&    $+2.9$&    $+0.99$& $+3.9$&         $+2.3$&          $+0.99$\\
                           &    $-3.2$&    $-1.4$&    $-0.98$& $-3.2$&         $-1.4$&          $-0.98$\\ \hline
\multirow{2}{*}{$\kappa_\mu(\%)$}&    $+20$&    $+5.4$&    $+2.6$& $+13$&         $+5.0$&          $+2.6$\\
                           &    $-23$&    $-5.0$&    $-2.7$& $-15$&         $-4.8$&          $-2.6$\\ \hline
\multirow{2}{*}{$\text{Br}_\text{inv}^{95\%}(\%)$}&    $+0.57$&    $+0.56$&    $+0.13$& $+0.56$&         $+0.56$&          $+0.13$\\
                           &    0
&    0
&    0
& 0&         0&          0\\ \hline
\multirow{2}{*}{$\text{Br}_\text{unt}^{95\%}(\%)$}&    $+23$&    
$+10$&    $+0.58$& $+9.0$&         $+7.5$&          $+0.59$\\
                           &    0&    0&    
0& 0&         0&          0\\ \hline \hline
\multirow{2}{*}{$\kappa_\Gamma(\%)$}&    $+28$&    
$+11$&    $+0.88$& $+9.6$&         $+7.9$&          $+0.87$\\
                           &    $-1.6$&    $-1.4$&    $-0.56$& $-1.5$&         $ -1.3$&          $-0.54$\\ \hline
\end{tabular}%
\caption{The $1\sigma$ upper and lower bound of the 11 parameters in the Higgs global fit under two different, forward coverage at 3 TeV MuC. The constraint values for $\text{Br}_\text{inv}$ and $\text{Br}_\text{unt}$ are the exclusion rate at 95\% C.L.
}
\label{Table:sensitivity_upper_lower_3TeV}
\end{table}

In this appendix, \autoref{Table:limit_3TeV_table} shows the result of a similar analysis at 3 TeV Muon Collider with an integrated luminosity of $1~\text{ab}^{-1}$. Note that 3 TeV Muon Collider performs better than 10 TeV under the scenario of $|\eta(\mu)|<4$. The reason comes from the different behavior of the signal process. The rapidity of forward muon at $\sqrt{s}=3$ TeV (shown in \autoref{fig:ForwardMuon_Eta_3_TeV}) is more central. Nevertheless, forward muon coverage can improve precision by more than 30\%, from 6.2\% to 3.9\%. We show the Higgs precision global fit results with 3~TeV muon collider in \autoref{Table:sensitivity_upper_lower_3TeV} for two forward muon detector coverages.
\clearpage

\bibliographystyle{utphys}
\bibliography{references}

\end{document}